\documentclass[10pt,journal]{IEEEtran}
\usepackage{algorithm}
\usepackage{algorithmicx}
\usepackage{algpseudocode}
\usepackage{amsmath,amsthm}
\usepackage{enumerate}
\usepackage{graphicx}
\usepackage{amsfonts}
\usepackage[subrefformat=parens,labelformat=parens]{subfig}
\usepackage{float}
\usepackage{graphicx}
\usepackage{xcolor}
\usepackage{hyperref}

\ifCLASSINFOpdf
\else
\fi
\makeatother

\begin{document}
%
\title{AF-DCGAN: Amplitude Feature Deep Convolutional GAN for Fingerprint Construction in Indoor Localization Systems}
%
%
%

\author{Qiyue Li,~\IEEEmembership{Member,~IEEE,}
        Heng Qu,
        Zhi Liu,~\IEEEmembership{Member,~IEEE,}
        Nana Zhou,
        Wei Sun,
        Stephan Sigg,~\IEEEmembership{Member,~IEEE,}
        and Jie Li
\thanks{Qiyue Li, Heng Qu, Nana Zhou and Wei Sun are with School of Electrical Engineering and Automation, Hefei University of Technology, Hefei, 230009, China. e-mail: liqiyue@mail.ustc.edu.cn, quhengedu@mail.hfut.edu.cn, nnzhou@mail.hfut.edu.cn and wsun@hfut.edu.cn.}
\thanks{Stephan Sigg is with  Department of Communications and Networking, Aalto University, Otakaari 5 a 02150 Espoo Finland; e-mail: stephan.sigg@aalto.fi}
\thanks{Zhi Liu is with Department of Mathematical and Systems Engineering, Shizuoka University, 5-627, 3-5-1 Johoku， Hamamatsu 432-8561, Japan; e-mail: liu@ieee.org}
\thanks{Jie Li (corresponding author) is with School of Computer and Information, Hefei University of Technology, Hefei, 23009, China. e-mail: lijie@hfut.edu.cn.}

\thanks{\textcircled{c} 20XX IEEE.  Personal use of this material is permitted.  Permission from IEEE must be obtained for all other uses, in any current or future media, including reprinting/republishing this material for advertising or promotional purposes, creating new collective works, for resale or redistribution to servers or lists, or reuse of any copyrighted component of this work in other works.}
}

{}

\maketitle

\begin{abstract}
With widely deployed WiFi network and the uniqueness feature (fingerprint) of wireless channel information, fingerprinting based WiFi positioning is currently the mainstream indoor positioning method, in which fingerprint database construction is crucial. However, for accuracy, this approach requires enough data to be sampled at many reference points, which consumes excessive efforts and time. In this paper, we collect Channel State Information (CSI) data at reference points by the method of device-free localization, then we convert collected CSI data into amplitude feature maps and extend the fingerprint database using the proposed Amplitude-Feature Deep Convolutional Generative Adversarial Network (AF-DCGAN) model. The use of AF-DCGAN accelerates convergence during the training phase, and substantially increases the diversity of the CSI amplitude feature map. The extended fingerprint database both reduces the human effort involved in fingerprint database construction and the accuracy of an indoor localization system, as demonstrated in the experiments.
\end{abstract}


%
\IEEEpeerreviewmaketitle

\section{Introduction} \label{intro}
As the demand for localization services increases, indoor localization technology based on fingerprint recognition has become the prevailing positioning technology due to its high precision and minimal hardware requirements. In addition to high accuracy, an indoor positioning system should have low complexity and require little processing time to accommodate mobile devices. Fingerprint-based indoor localization is an effective method that can satisfy these requirements; however, the received signal strength (RSS) or channel state information (CSI) from surrounding access points must be measured at each reference point to build a fingerprint database \cite{3}.

Most of the fingerprint-based localization systems are based on wireless local area networks (WLANs), which are available in public places \cite{8353839}, \cite{8326317}. Fingerprint-based localization consists of two basic phases: 1) an offline phase (collect data  at each reference point to construct the fingerprint data and train the classification model for the online phase) and 2) an online phase (receive data online, compare the received data with the fingerprint database to achieve localization) \cite{4}. The training phase is used to construct the fingerprint database by collecting and preprocessing survey data related to each reference point's position. During the online phase, a mobile device records real-time data and compares the received data with the database. The reference point that most closely matches the received data is assumed to be device's location.

Many existing indoor localization systems use RSS for fingerprints due to simplicity and low hardware requirements. For example, the \textit{Horus} system uses a probabilistic method to estimate location with RSS data \cite{5}. However, RSS data has high variability over a fixed location due to multipath effects in indoor environments \cite{6}. This high variability can introduce significant localization errors. Additionally, RSS values consist of relatively coarse information that does not fully exploit the many subcarriers in an orthogonal frequency-division multiplexing (OFDM) system. Instead, in the widely used OFDM systems, CSI provides more precise multipath information than RSS does by exploiting the different signal strengths and phases in different subcarriers \cite{8304587}. Some commercial off-the-shelf network interface cards with IEEE 802.11n standard provide detailed subcarrier amplitude and phase information in the form of CSI. Thus, this paper considers the CSI-based fingerprint for indoor localization. 

One major issue of fingerprint-based localization is that it is difficult to determine how much data is needed to obtain the desired accuracy during the offline phase. In the next section, we show that sampling more fingerprint data provides better results. Thus, building a sufficiently large fingerprint database is vital to high localization performance. However, this task is time consuming and labor-intensive \cite{3}. As an example, it takes half an hour to collect the fingerprint data at one reference point in our experiments. This excessive time to perform fingerprint data collection affects the popularity and applications of fingerprint-based localization. 

To reduce the data collection cost, several methods have been proposed \cite{3}, \cite{7}, \cite{8}. In particular, in \cite{3}, the authors propose a method based on compressive sensing to recover absent fingerprints. Their approach shows the hidden structure and redundancy characteristics of fingerprints in a merging matrix. In \cite{7}, the authors present a semi-supervised manifold learning technique for building a fingerprint database from partially labeled data, where only a small portion of the signal strength measurements must be marked with the corresponding coordinates. Note that these methods only reduce the number of reference points or try to recover fingerprints; they fail to reduce the number of samples to be collected and still require a significant amount of time for data collection.

To expand the fingerprint database while reducing the human effort, we present a method to increase the amount of training data collected at each reference point based on generative adversarial nets (GAN) \cite{goodfellow2014generative}, \cite{cao2019recent}. Specifically, we first transform the CSI data collected at each reference point into amplitude feature maps. Then, through pixel transformation, each amplitude feature map is transformed into an image with the same resolution to construct the initial fingerprint database. Because the initial database is constructed by mapping the amplitude feature maps of the reference points, it contains the location information of all reference points. Then, we propose an Amplitude-Feature Deep Convolutional Generative Adversarial Network (AF-DCGAN) model as well as a corresponding training algorithm that generates images similar to the original amplitude feature maps. Finally, the generated amplitude feature maps are merged into the initial fingerprint database, resulting in an expanded database. As described in the following section, the localization accuracy improves as the number of samples in the fingerprint database increases. We also evaluate the proposed fingerprint construction method through extensive experiments in a typical indoor classroom environment. The accuracy of the initial database reaches 1.34~m, while the accuracy of the expanded database reaches 0.92~m, indicating the effectiveness of the proposed method.

The main contributions of this paper are as follows:
\begin{enumerate}[1.]

\item We build a fingerprint database by converting processed CSI data into amplitude feature maps. This approach visualizes the position of the sampling points and allows us to visually determine the locations of the testing points.

\item Based on the nature of the amplitude feature maps, an AF-DCGAN model is proposed that converges quickly and generates samples with improved diversity. We use the AF-DCGAN model to generate additional amplitude feature maps of the sampling points' positions, which reduces the collection time associated with each single sample point and saves human effort.

\item Extensive experiments are conducted and the results show that the proposed scheme can provide better performance compared to state-of-the-art schemes.

\end{enumerate}

The remainder of this paper is organized as follows. Section~\ref{related_work} reviews the related work. Section~\ref{basic_idea} briefly describes the basic idea underlying this paper. Section~\ref{csi_collection_feature_map} introduces the CSI and the amplitude feature map conversion technique. Section~\ref{af_dcgan_model_method} presents the method for expanding the amplitude feature maps based on AF-DCGAN, including the training algorithm and data-generating steps. Simulation and experimental results are shown in Section~\ref{experiment}. Finally, Section~\ref{conclusion} concludes the paper.

\section{Related work} \label{related_work}

WiFi localization technology can be divided into two categories: positioning based on propagation models and positioning based on fingerprint. Although the propagation model-based methods do not require signal sampling, fingerprint-based positioning achieves higher accuracy \cite{9}, \cite{8307353}, \cite{11}, \cite{12}. Therefore, WiFi indoor positioning based on fingerprints is gaining popularity.

Although building a fingerprint database for localization purposes is highly efficient, collecting the data to construct the fingerprint database usually requires significant human effort. Many researchers have proposed solutions for the construction of the fingerprint database to reduce human effort. In \cite{3} and \cite{7857072}, a novel approach based on compressive sensing is presented to recover absent fingerprints. Jun \emph{et al.} \cite{16} present the design, implementation, and evaluation of AP-Sequence. This fingerprint-based localization system achieves extremely low overhead in fingerprint map construction and maintenance. The method in \cite{17} leverages a more stable RSSI gradient to build a gradient-based fingerprint map by comparing the absolute RSSI values at nearby positions. A novel fingerprint collection technique is proposed in \cite{18} that detects WiFi APs to form WiFi fingerprints from the signals collected by ZigBee interfaces. In \cite{19}, an FM-based indoor localization system that does not require proactive site profiling is presented to construct the fingerprint database based solely on an estimate of indoor RSS distribution.

The authors of \cite{7980032} propose a fingerprint-based device-free localization system named \textit{iUpdater} to significantly reduce the labor cost and increase the accuracy. It is able to accurately update the whole database with RSS measurements at a small number of reference locations, thus reducing the human labor cost. Milioris \emph{et al.} \cite{21} use the Matrix completion framework to build complete training maps from a part of the reference fingerprints by learning the relevant fingerprint structure. In \cite{8003484}, the authors propose \textit{AcMu}, an automatic and continuous radio map self-updating service for wireless indoor localization that exploits the static behaviors of mobile devices. By accurately pinpointing mobile devices with a novel trajectory matching  algorithm, they use stationary mobile devices as reference points to collect real-time RSS samples. The authors of \cite{23} propose the Enriched Training Database (ETD), which is a web-service that enables the management and storage of training fingerprints and includes additional “enriching” functionality. The user can automatically generate virtual fingerprints based on propagation modeling of the virtual training points through the enriching functionality. The same authors also propose a new method to acquire training fingerprint locations that eliminates the burden of manually defining training points and covers areas with insufficient density to train fingerprints. In \cite{24}, the authors propose a novel method to construct a comprehensive fingerprint database by using the radio propagation model. To address the limitations of indoor dynamic measurements, the authors in \cite{25} propose a Gaussian process regression for fingerprint-based localization that uses realistic and virtual indoor dynamic measurement data.

However, most of these methods use modeling to reduce the number of reference points or recover fingerprints based on existing partial fingerprints. The accuracy of localization systems decreases dramatically when many reference points are omitted to reduce labor cost. In this paper, we attempt to sample data at all the reference points and then generate additional data using an AF-DCGAN model, which effectively expands the number of fingerprints in the database. Thus, we obtain better positioning performance with the expanded fingerprint database, while also reducing the human effort.

\section{Basic idea} \label{basic_idea}

\begin{figure}[htb]
\centering
\subfloat[The receiving antennas (MP) are placed behind the measurement area to cover all the experiment area.]{
  \includegraphics[width=2.5in]{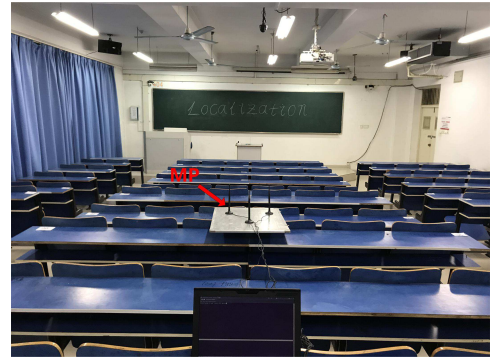}\label{fig_deployment_mp}}\hfill
\subfloat[The router is placed in front of the measurement area.]{
  \includegraphics[width=2.5in]{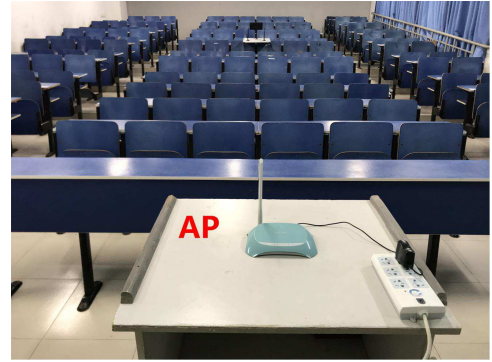}\label{fig_deployment_ap}}\hfill
\subfloat[Floorplan of the experiment area.]{
  \includegraphics[width=2.5in]{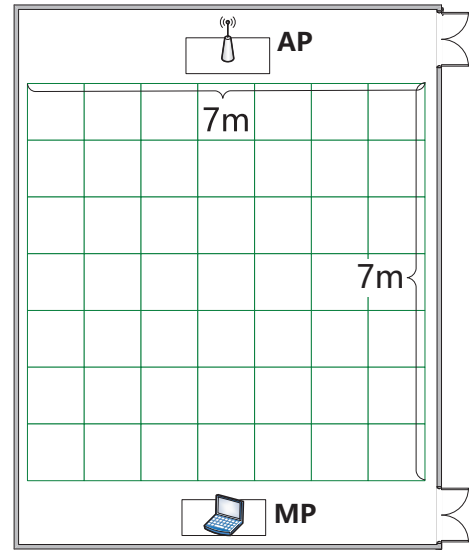}\label{fig_deployment_floorplan}}\hfill
\caption{Illustration of the experimental enviroment (a classrom). The router is placed at the front of the measurement area. The receiving antennas are placed behind the measurement area to cover the measurement area.} \label{fig_experiment_environment}
\end{figure}

In a fingerprint-based indoor localization system, the more data the fingerprint database contains, the higher its positioning accuracy is. We first conduct experiments to show this relationship. We use different methods to construct fingerprint databases and then perform localization experiments in a classroom. The environment and floor plan are illustrated in Fig.~\ref{fig_experiment_environment}. We deploy a TL-WR742N wireless router as the transmitter, equipped with one transmit antenna, and a ThinkPad laptop equipped with Intel Wireless Link 5300 NICs (IWL5300) with three receive antennas. We divide the classroom into 49 $(7\times 7)$ sections and set the center point in each section as the corresponding reference point.

\begin{figure}[htb]
\centering
\includegraphics[width=3.5in]{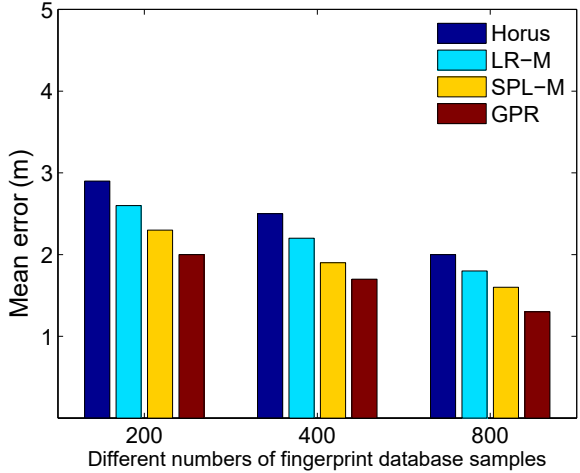}
\caption{Localization accuracy of different fingerprint database sizes. As the number of samples increases, the mean localization error gradually decreases.}
\label{fig_localization_error_database_size}
\end{figure}

At each reference point, we collect 800 samples and test several different fingerprint database construction methods, including the Horus method \cite{5}, Gaussian process regression (GPR) \cite{26}, Low-Rank matrix fill (LR-M) method \cite{27}  and the thin spline interpolation method (SPL-M) \cite{28}. The Horus system requires reference point data uses location-clustering techniques to reduce the computation requirements of the algorithm. The GPR method uses half the reference points and infers the posterior received RSS mean and variance at other points to build a fingerprint database. The LR-M represents the distribution of wireless fingerprints as a low-rank matrix and constructs a dense radio map from relatively sparse measurements using a revised low-rank matrix completion method. The SPL-M method explores the use of different interpolation functions to complete the fingerprint mapping necessary to achieve the required accuracy. During the experiments, we use 200, 400 and 800 samples at each reference point to build three fingerprint databases of different sizes. Then, we use a k-means clustering algorithm to perform positioning. The results are shown in Fig.~\ref{fig_localization_error_database_size}. The Y-axis indicates the mean error of localization. These experiments show that, as the size of the fingerprint database for each sampled data location increases, the localization accuracy gradually increases. Larger training datasets achieve more satisfactory results than smaller training datasets. As the number of samples used in the database increases, the mean error decreases.

However, it requires tremendous labor to sample a large amount of wireless fingerprints. Inspired by unsupervised learning \cite{Shrivastava_2017_CVPR}, we use GAN to generate more sample data for each reference point, to expand the fingerprint database and to improve the accuracy of the localization system. In the following sections, we convert the CSI data into amplitude feature maps and then extend the fingerprint database by the proposed AF-DCGAN model. With this model, the convergence process in the training phase is accelerated and the diversity of the generated CSI amplitude feature map is increased dramatically. Based on the extended fingerprint database, the accuracy of the indoor localization system can be improved with reduced human effort.

\section{CSI collection and feature map conversion} \label{csi_collection_feature_map}
In the field of wireless communication, Channel State Information (CSI) is a channel attribute of the communication link. It describes the signal’s attenuation factor on each transmission path, that is, the value of each element in the channel gain matrix $H$, such as signal scattering, multipath fading, shadow fading or power decay of distance.

In the widely used OFDM system, CSI provides more multipath information than does RSS, by including the signal strength and phase of different subcarriers. Recently, some IEEE 802.11n standard commercial off-the-shelf network interface cards provide access to detailed subcarrier amplitude and phase information via CSI. Specifically, with the Intel 5300 NIC, a sample of Channel Frequency Response (CFR) over WiFi bandwidth can be obtained as CSI information, including the number of transmit antennas $N_t$, the number of receive antennas $N_r$, the number of subcarriers $N_S$, the packet transmission frequency $f$ and the CSI matrix $H$ \cite{29}, which is a $N_t\times N_r\times N_S$ imaginary number matrix as follows:

\begin{equation}
	H=(H_{uv})_{N_t\times N_r}
\end{equation}

Each pair of transmit-receive antennas (a TX-RX pair) is a link, and $H_{uv}$ is the CSI data of the link formed by TX $u$ and RX $v$, containing the information of $N_S$ subcarriers.

\begin{equation}
\begin{split}
	& H_{uv}=(h_1^{uv},\dots,h_k^{uv},\dots, h_{N_s}^{uv})^T, \\
	& 1 \leq  u \leq  N_t, 1 \leq v \leq  N_r, 1 < k < N_S
\end{split}
\end{equation}

Each $h_k$ characterizes the amplitude and phase of the corresponding subcarrier which can be expressed as
\begin{equation}
h_k^{uv}=\left | h_k^{uv} \right |e^{j\angle h_k^{uv}}, 1 \leq k \leq N_S
\end{equation}

where $\left | h_k^{uv} \right |$ denotes the amplitude response and $e^{j\angle h_k^{uv}}$ denotes the phase response of subcarrier $k$. It has been shown that the raw phase values reported by the wireless network card are inaccurate, and the phase values vary greatly in some frequency bands even in static environments \cite{Xie2015}. Thus, currently in most CSI-based localization algorithms the phase values are not used to determine the positions of the target. In contrast, the amplitude value for a given sub-carrier maintains a good stability at a certain location, which is thus adapted in CSI-based localization approaches \cite{XIAO201773}. To show the superiority of our algorithmic approach, and to directly compare with related work, we also limit ourselves to use amplitude information only. Generalization to phase information is left for future work.

\begin{figure}[htb]
\centering
\subfloat[]{
  \includegraphics[width=3in]{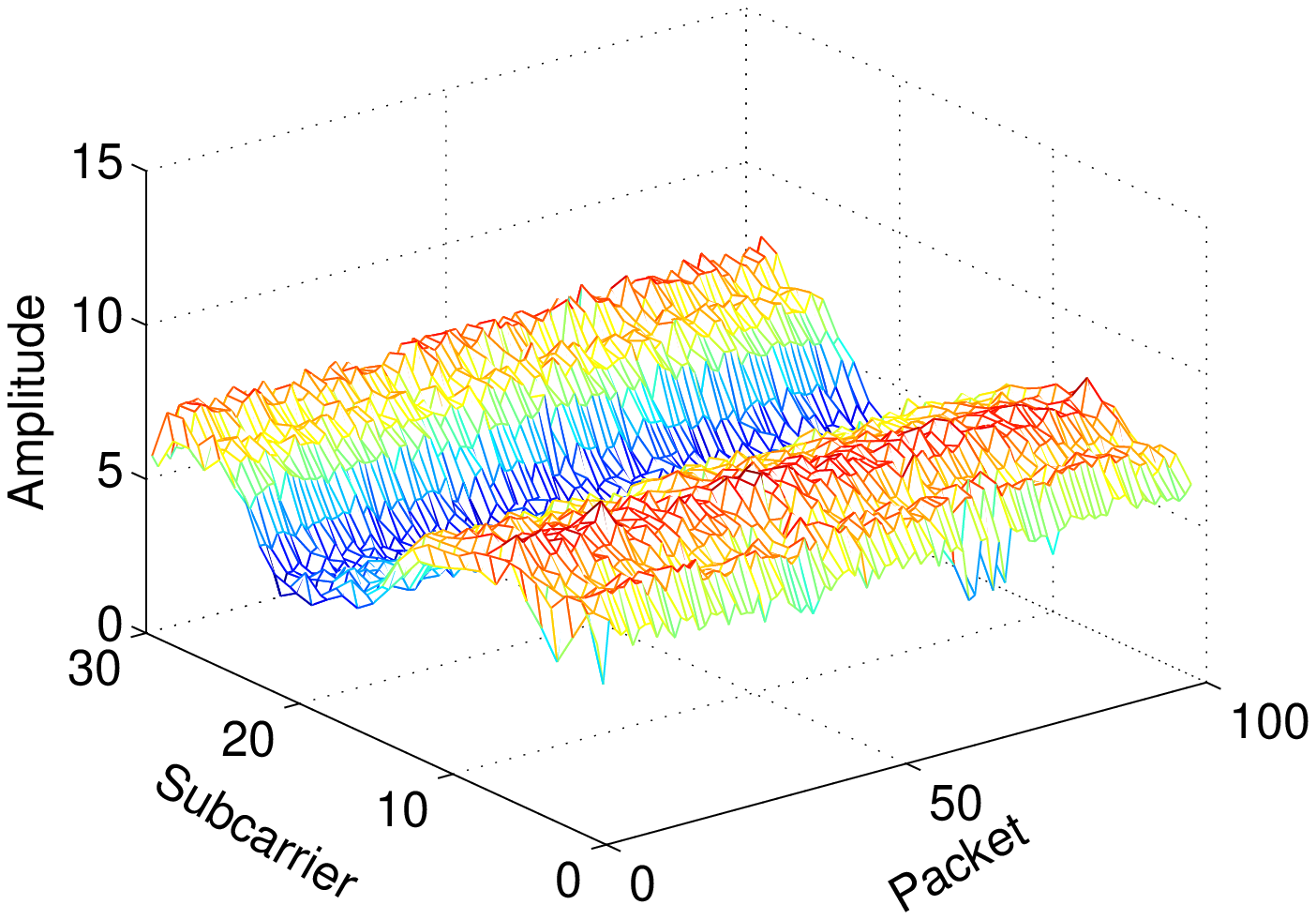}\label{fig_csi_packets_a}}\hfill
\subfloat[]{
  \includegraphics[width=3in]{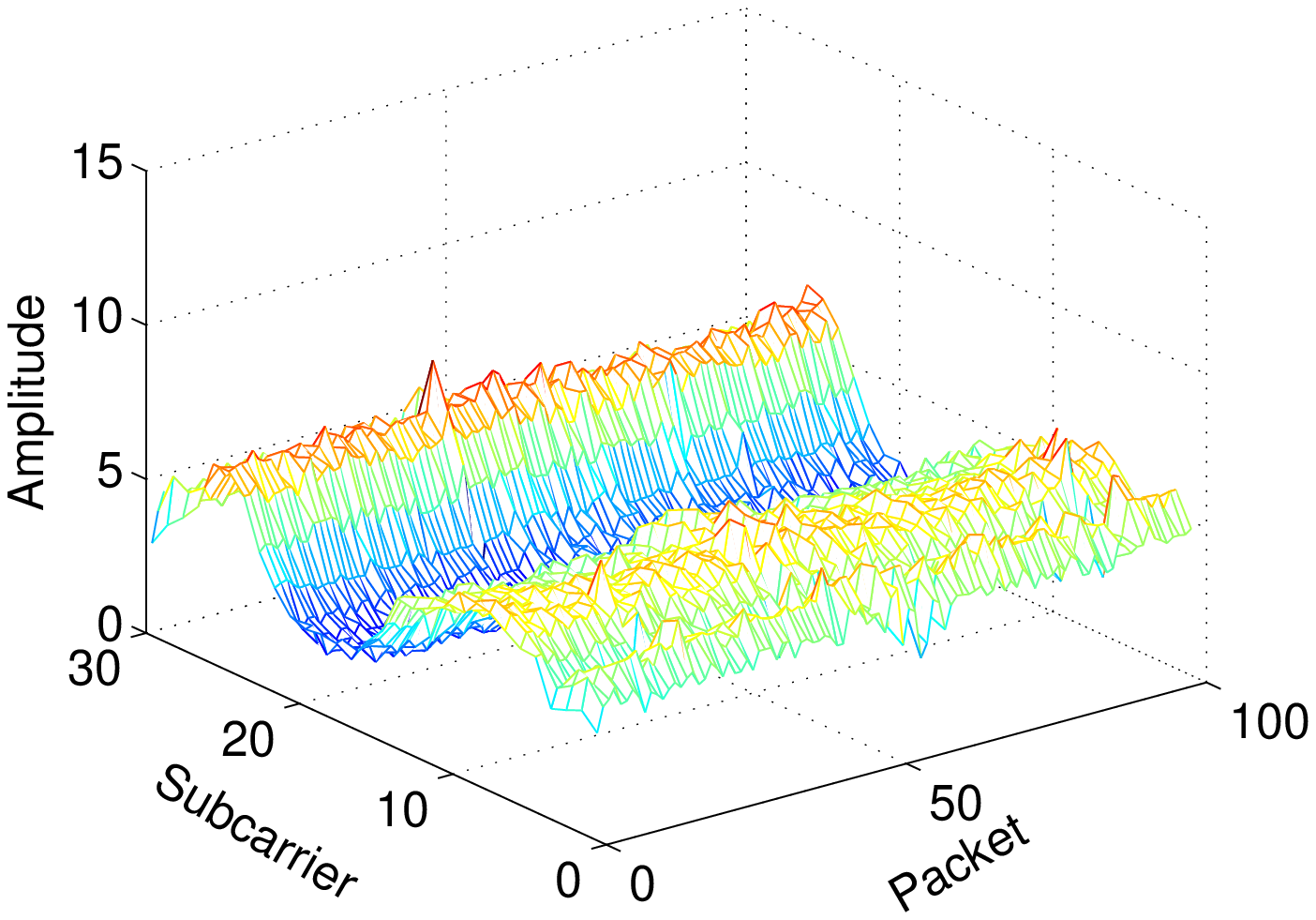}\label{fig_csi_packets_b}}\hfill
\subfloat[]{
	\includegraphics[width=3in]{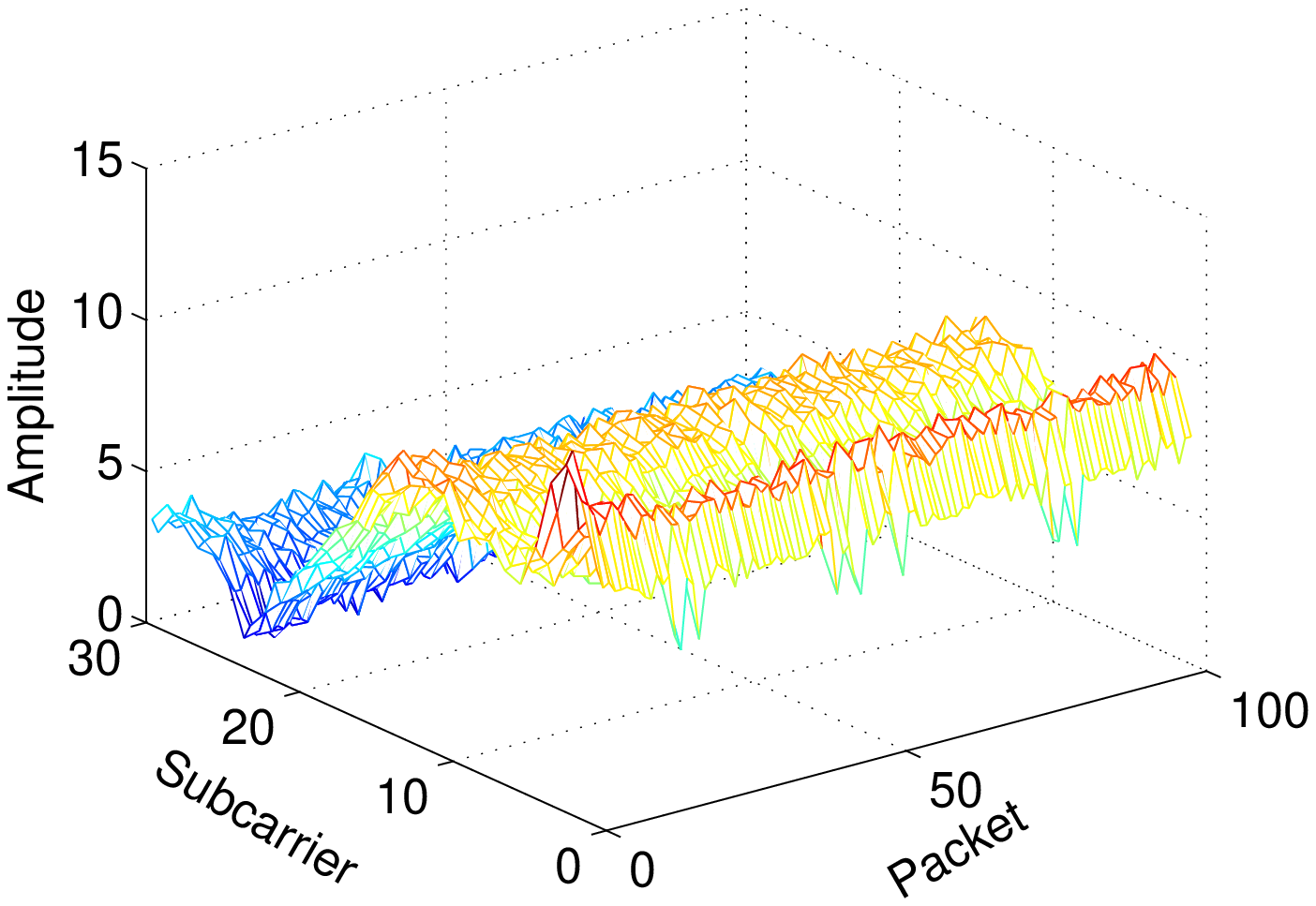}\label{fig_csi_packets_c}}\hfill
\caption{ Large changes occur in CSI amplitudes at different locations: (a), (b) and (c) show three-dimensional patterns of amplitude, subcarriers and packets at three positions separated by 1 m and 5m.} \label{fig_csi_packets}
\end{figure}

In a defined space as shown in Fig.~\ref{fig_experiment_environment}, the CSI data clearly differs when people stand in different places. Fig.~\ref{fig_csi_packets} shows the amplitude ($\left | h_k^{uv} \right |$) of all subcarriers in one link over time/packets when a person stands at three different positions in a classroom. Fig.~\subref*{fig_csi_packets_a} is the three-dimensional amplitude  map of the reference position, and the distance of the position of Fig.~\subref*{fig_csi_packets_b} and Fig.~\subref*{fig_csi_packets_c}  from Fig.~\subref*{fig_csi_packets_a} is 1m and 5m, respectively. As the figure shows, the difference in the amplitude of the CSI increases as the distance between the measurement locations increases. These characteristics of CSI amplitude changes are exploited in building a fingerprint database to perform indoor localization.

\subsection{CSI collection} \label{csi_collection}
First, we evenly divide the designated indoor space into $M$ sampling spaces and use the centers of the sampling spaces as reference points to form a reference point (RP) set as follows.
\begin{equation}
	RP=[RP_1, \cdots, RP_i, \cdots, RP_M]
\end{equation}
where $RP_i$ denotes the reference point in the $i$-th ($1 < i < M)$ square grid.

Assume $N_{ap}$ wireless access points (wireless routers in IEEE 802.11) are deployed in the indoor space. Then, each CSI sample has $N_{ap}\times N_t\times N_r \times N_S$ dimensions. During the collection stage, we obtain $N_X$ samples of CSI data at a fixed rate when people stand at different reference points, forming a time series set of the $i$-th reference point as follows.

\begin{equation}
	CSI_{N_X}^{i}=\{csi_{1,N_X} ^i,\cdots, csi_{m,N_X}^i,\cdots, csi_{N_t\times N_r,N_X}^i\}
\end{equation}
where $csi_{m,N_X}^i$ denotes the $N_X$ WiFi signals of the $m$-th ($1 < m <N_t\times N_r$) link received by the $i$-th reference point, and $csi_{m,N_X}^i$ is a two-dimensional imaginary number matrix with size $N_X\times N_S$.

\subsection{Convert CSI to amplitude feature maps}
The CSI amplitude varies significantly in different positions, which can be exploited to perform indoor localization. To obtain a fingerprint corresponding to a specific location, we represent the characteristics of that location by plotting the amplitude of the CSI as a feature map.

We randomly select 100 out of $N_X$ rows of the two-dimensional imaginary number matrix $csi_{m,N_x}^i$ for $m=1\cdots T$ to form $T$ two-dimensional matrices of $100\times N_S$, to reconstruct the position information set at the $i$-th reference point $CSI_i^{'}$ as follows.
\begin{equation}
	CSI_i^{'}=\{csi_{1,T}^{'i},\cdots,csi_{m,T}^{'i},\cdots, csi_{N_t\times N_r,T}^{'i}\}
\end{equation}
where $csi_{m,T}^{'i}$ denotes $T$ imaginary number matrices of $100\times N_S$ of the $m$-th link at the $i$-th reference point $RP_i$.

The real and imaginary parts of the imaginary number matrices in the reconstructed position information set $CSI_i^{'}$ are selected to create amplitude feature maps to obtain the amplitude feature maps set $\Phi_i$ of $N_t\times N_r$ links at $RP_i$. Further, we obtain a set of $\Phi$ of amplitude feature maps at $M$ reference points.

\begin{equation}
	\Phi_i=\{\phi_{1,n}^{i} \cup \cdots\cup \phi_{k,n}^{i} \cup \cdots \cup \phi_{N_t\times N_r,n}^{i}\}
\end{equation}
\begin{equation}
		\Phi=\{ \Phi_1,\cdots, \Phi_i, \cdots, \Phi_M\}
\end{equation}
where $\phi_{k,n}^{i}$ denotes $n$ amplitude feature maps of the $k$-th link at the $i$-th reference point, and $\Phi_i$ denotes $n$ amplitude feature maps of $N_t\times N_r$ links at the $i$-th reference point.

\begin{figure}[htb]
\centering
\subfloat[]{
  \includegraphics[width=3.5in]{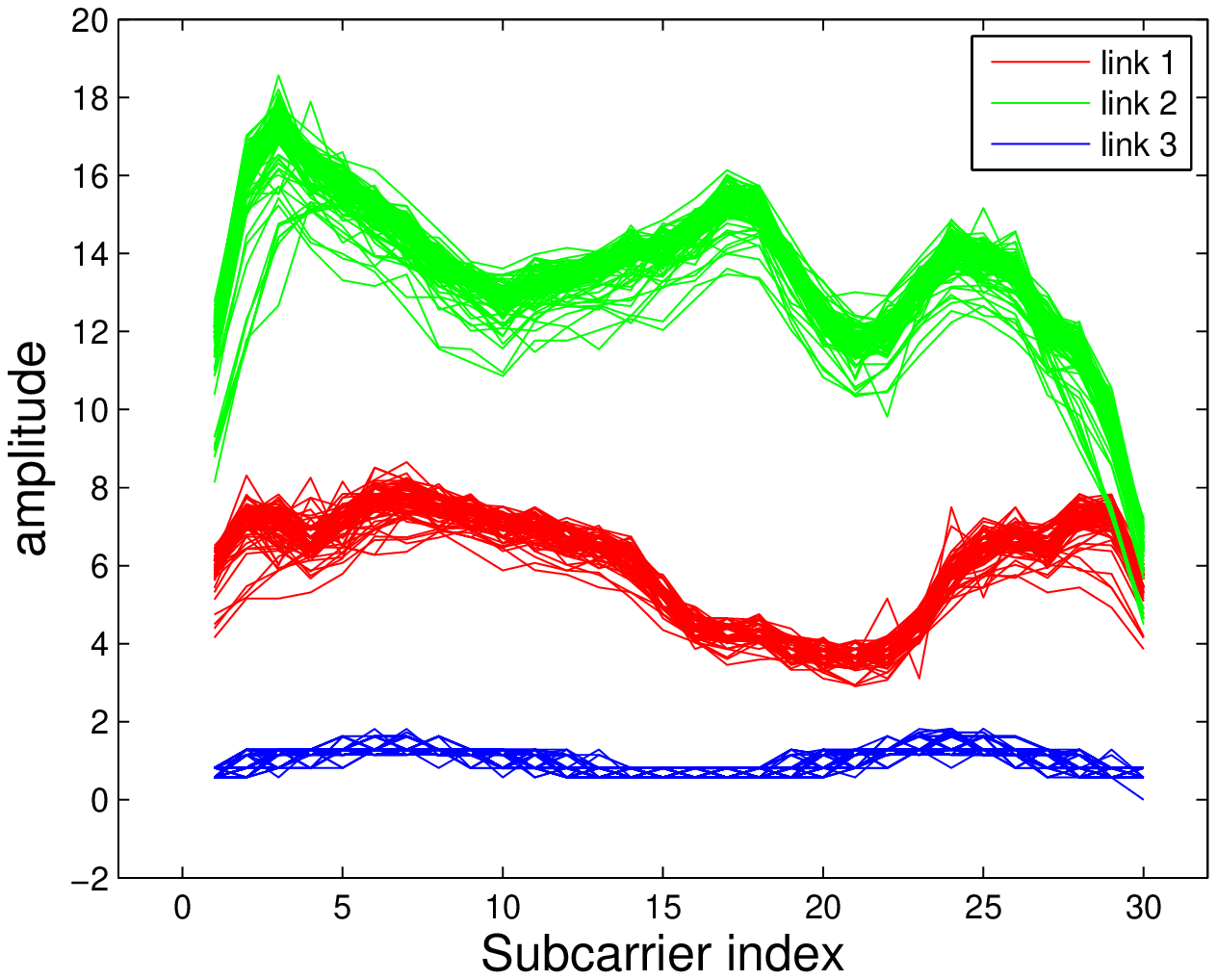}\label{fig_afm_subcarrier_a}}\hfill
\subfloat[]{
  \includegraphics[width=3.5in]{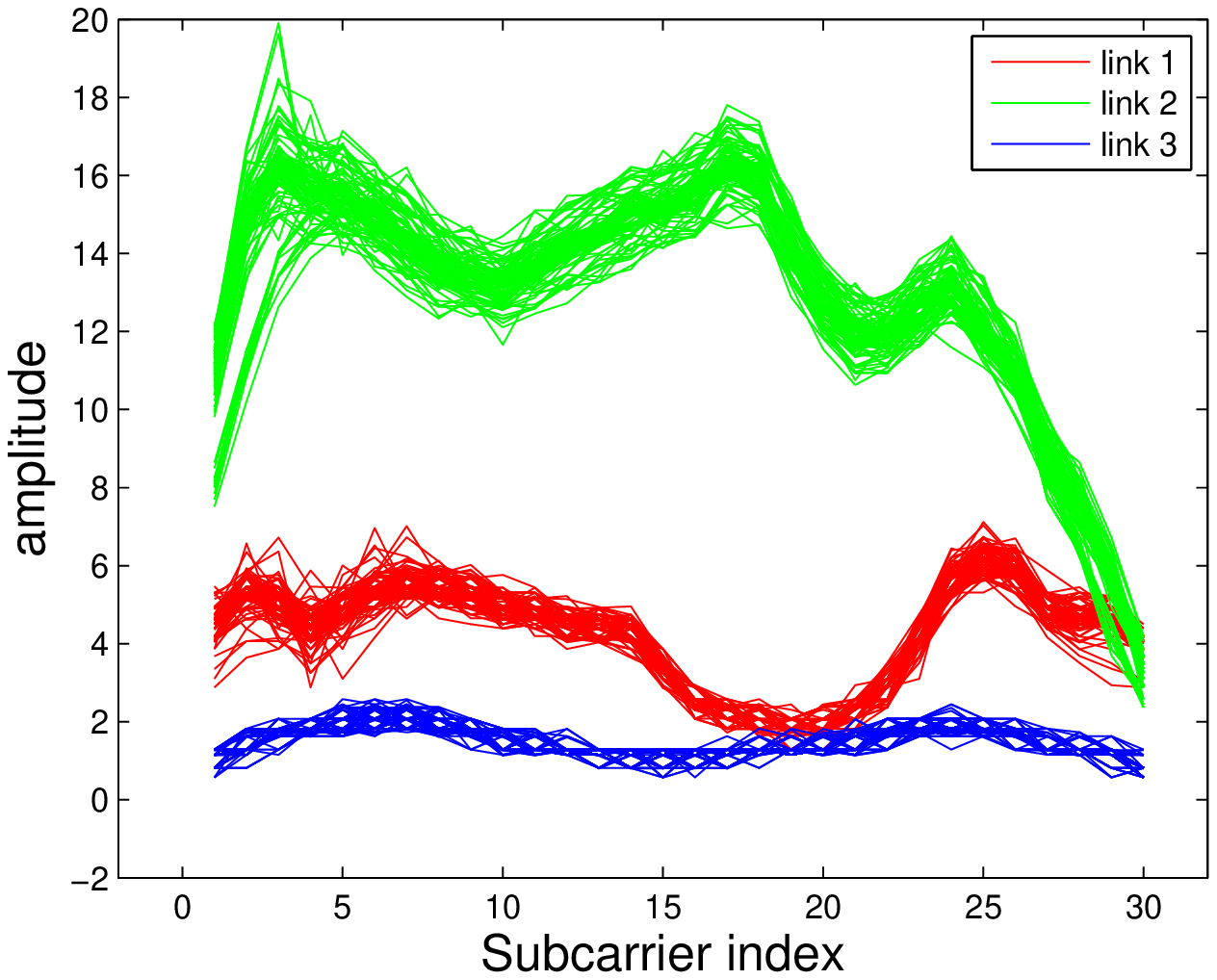}\label{fig_afm_subcarrier_b}}\hfill
\caption{Different colors represent different links: (a) and (b) are feature maps of two sampling points from the $M$ sampling points. Given three antennas at the receiver and one antenna at the transmitter, each link represents a path between the transmitting end and the receiving end. Each subcarrier of these three links has different amplitudes, thus resulting three lines in the figures.} \label{fig_afm_subcarrier}
\end{figure}

As shown in Fig.~\ref{fig_afm_subcarrier}, the amplitude feature maps of two positions in $M$ reference points obtained through data processing are created so that the position information indicated by the CSI data can be visualized.

In order to adapt to the image classification model and to improve the accuracy of image classification, we transform the data in the $\Phi_i$ of $RP_i$ into a picture with a resolution of $256\times 256$. A training set of the $i$-th reference point is obtained as follows. Then, we identify the initial fingerprint database $\Phi^{'}$.

\begin{equation}
	\Phi_i^{'}=\{\phi_{1,n}^{'i}\cup \cdots\cup \phi_{k,n}^{'i} \cup \cdots \cup \phi_{N_t\times N_r,n}^{'i}\}
\end{equation}
\begin{equation}
		\Phi^{'}=\{ \Phi_1^{'},\cdots, \Phi_i^{'}, \cdots, \Phi_M^{'}\}
\end{equation}
where $\phi_{k,n}^{'i}$ denotes the $n$ amplitude feature maps of the $k$-th link at $RP_i$ after the resolution transform, and $\Phi_i^{'}$ denotes the $n$ amplitude feature maps of the $N_t\times N_r$ links at $RP_i$.

\section{AF-DCGAN model and training method} \label{af_dcgan_model_method}

To reduce the sampling time and human effort, we use GAN to generate additional amplitude feature maps for each reference point, thus expanding the initial fingerprint database. In this way, we obtain additional samples similar to the initial fingerprint database to expand the fingerprint database and to improve the localization accuracy without additional human effort.

\subsection{AF-DCGAN model} \label{af_dcgan_model}
GAN is inspired by two-player game theory. The two players in the GAN model are the generative model ($G$) and discriminative model ($D$) \cite{30}. $G$ captures the distribution of sample data to generate a sample similar to the training data with added noise that obeys a certain distribution (uniform, Gaussian, etc.). Using this approach, the generated samples approximate real samples taken at the same location. $D$ is a binary classifier that estimates the probability that a sample comes from the training data. If the sample comes from the real training data, $D$ outputs a high probability; otherwise it outputs a small probability.

During training, one side is fixed, and the network weights at the other side are updated and alternately iterated. During the training process, both sides attempt to optimize their networks, forming a rivalry that continues until the two parties reach a dynamic balance (Nash Equilibrium). $G$ restores    the distribution of the training data and creates samples increasingly similar to the real data until $D$ can no longer discriminate the results at an accuracy above 50\%, at which point the discriminator and generator have reached Nash equilibrium. $D$ and $G$ play the following two-player minimax game with the value function $V(D,G)$ as follows:

\begin{equation}
\begin{split}
	\min_{G} \max_{D}V(D,G)= & E_{x\sim P_{data}(x)}\{logD(x)\}+ \\
	& E_{z\sim P_z(z)}\{log(1-D(G(z)))\}
\end{split}
\end{equation}
where $P_{data}(x)$ denotes the real sample set, $z$ denotes the signal with a uniform distribution. $P_z(z)$ denotes the fake sample set, $G(z)$  is the output of the generator, and $D(x)$ is the output of the discriminator.

Convolutional neural networks perform well at supervised learning tasks but have been rarely used for unsupervised learning. DCGAN \cite{31} combines a CNN using supervised learning with a GAN using unsupervised learning. In this paper, we extend the fingerprint database by using DCGAN to generate additional amplitude feature maps. DCGAN is a well-established GAN model; however, in our application, it converges slowly when generating amplitude feature maps and the model collapses during the training process. To resolve these problems, the AF-DCGAN model is proposed in the following section. Using AF-DCGAN accelerates convergence and increases the diversity of the generated CSI amplitude feature maps dramatically.

\begin{figure*}[htb]
\centering
\subfloat[]{
  \includegraphics[width=7in]{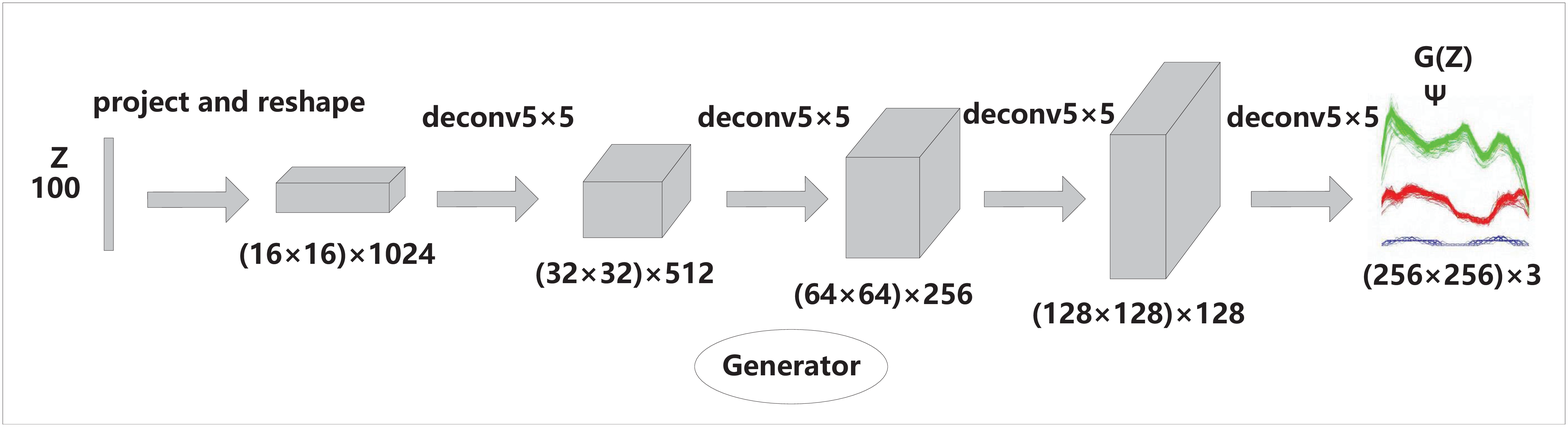}\label{fig_generative_model}}\hfill
\subfloat[]{
  \includegraphics[width=7in]{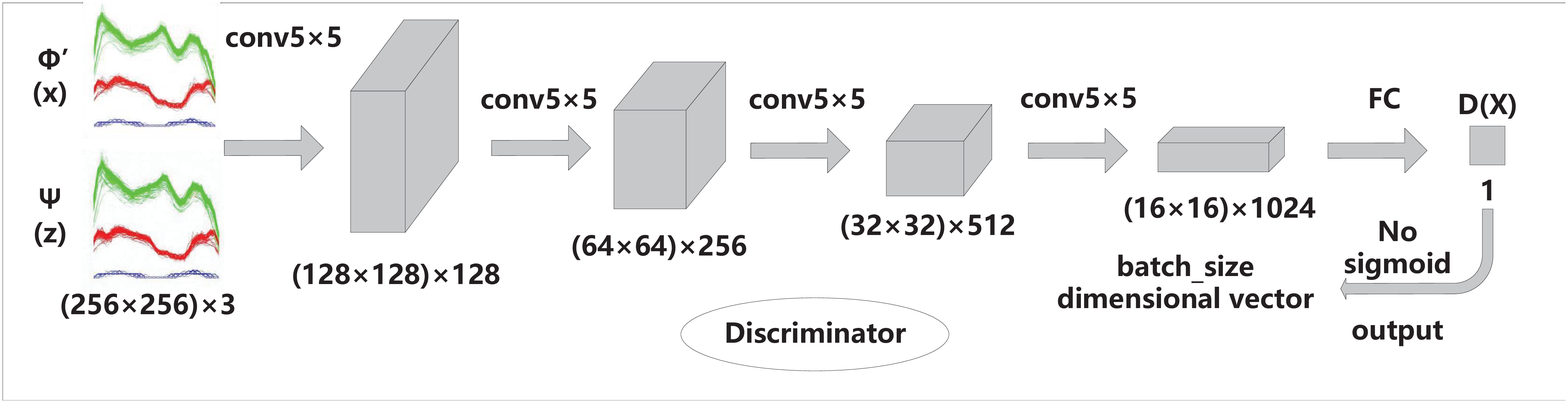}\label{fig_discriminative_model}}\hfill
\caption{Network structure of the AF-DCGAN model: (a) $G$ is the generative model, implemented by a deconvolutional network; $Z$ is the signal, which has a uniform distribution; $deconv$ represents the deconvolution layer in the CNN model; $\Psi$ denotes the feature maps generated by the deconvolutional layers in the generator. (b) Discriminative model, implemented by a convolutional network. $\Phi'$ are feature maps from the training set; $\Psi$ are the generated samples; $conv$ represents the convolution layer in the CNN model; $D(x)$ indicates the probability that the input sample stemmed from the training set corresponding to $x$ or $G(z)$ as the input; $FC$ denotes the fully connected layers in the CNN models.}
\end{figure*}

The AF-DCGAN model is depicted in Fig.~\subref*{fig_generative_model} and Fig.~\subref*{fig_discriminative_model}. And it is worth noting that in AF-DCGAN, the final deconvolution layer size of the generator is $128\times 128$, which allows us to directly generate images with $256\times 256$ pixels.

\subsection{Training process}
\begin{figure}[htb]
\centering
\subfloat[]{
  \includegraphics[width=3in]{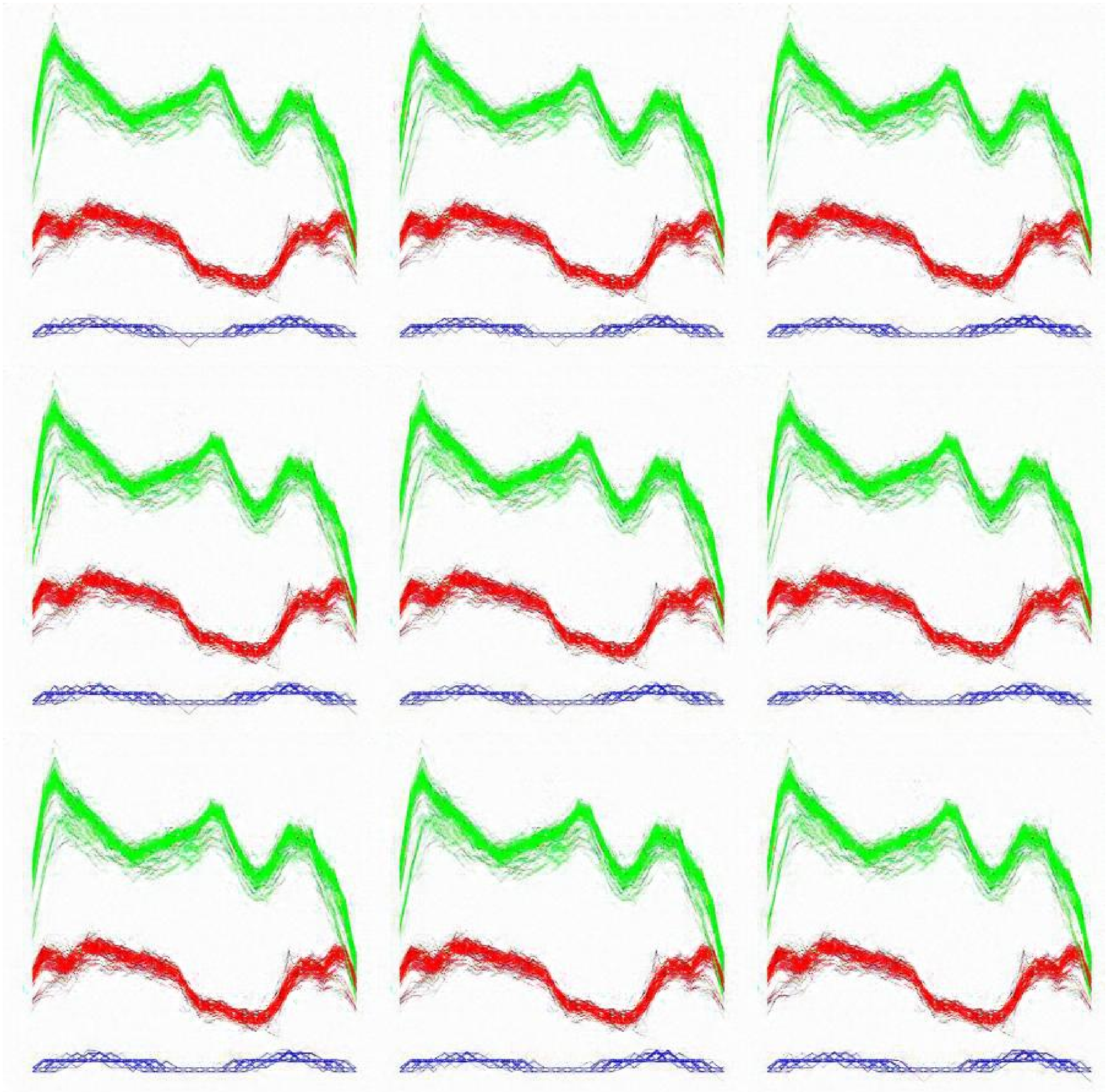}\label{fig_afm_dcgan}}\hfill
\subfloat[]{
  \includegraphics[width=3in]{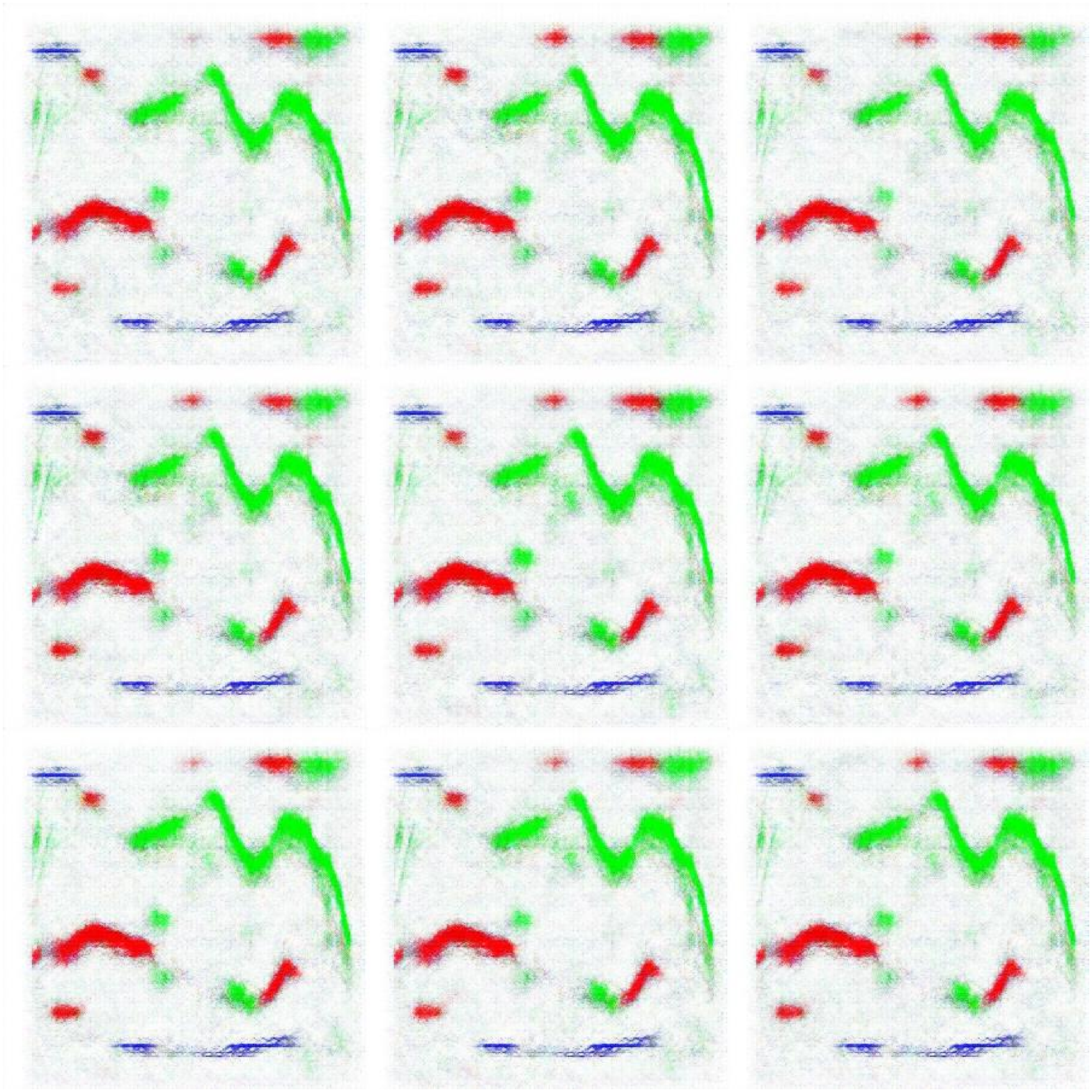}\label{fig_afm_wgan}}\hfill
\caption{CSI amplitude feature maps generated by GAN model: (a) Maps from DCGAN in the 100th epoch. It can be seen from the figure that all the subgraphs are exactly the same and the sample diversity is very poor. (b) Maps from WGAN in the 100th epoch. The model is difficult to converge during the training process.} \label{fig_afm_dcgan_wgan}
\end{figure}

From WGAN \cite{32}, we know that the GAN has problems such as difficult training, loss of generators and discriminators, lack of diversity of training samples and in generating training samples. Therefore, generating a large number of CSI amplitude feature maps by GAN is not easy. We test the performances of WGAN and DCGAN, and the results are shown in Fig.~\ref{fig_afm_dcgan_wgan}.

When the traditional DCGAN is applied to expand the CSI fingerprint database, the diversity of the generated amplitude feature maps is poor and yields no performance gains for the indoor localization. Although WGAN reduces the difficulty of GAN training, convergence is not reached in some settings, and the generated pictures are worse than those of DCGAN. When WGAN is used for our amplitude feature maps, the model collapses quickly during the training process, and the generated data seems to be random, as shown in Fig.~\subref*{fig_afm_wgan}.

WGAN proposes to use the Wassertein distance as an optimization method to train GAN, but this still leaves difference between the mathematical and real implementation. The Wassertein distance needs to satisfy the strong continuity condition: Lipschitz continuity. To satisfy this condition, the authors forced the Lipschitz continuity to be met by limiting the weight to a range, but it also creates hidden dangers. The Lipschitz constraint is in the sample space, and the discriminator function $D(x)$ gradient value is required to be no more than a finite constant $K$. The weighted value constraint ensures the boundedness of the weight parameter and indirectly limits the gradient information. To solve these problems, we imitate part of WGAN to improve DCGAN in AF-DCGAN to improve the diversity of the generated samples. We remove the Sigmoid function from the last layer of the discriminator and return the normalized fully connected layer to make sure the value of $D(x)$ in loss function locates with [0,1]. Specifically, after sending the first batch of training data into discriminator, we normalize the output of full connection layer  $D(x)$ with function $D_{norm}(x)=\frac{D(x)-D_{1b\_min}(x)} {D_{1b\_max}(x)-D_{1b\_min}(x)}$, where $D(x)$ is the output value of full connection layer, $D_{1b\_min}(x)$ and $D_{1b\_max}(x)$ are the minimum and maximum output value in the first batch of training data respectively. The other batches are normalized in the similar way with the same $D_{1b\_min}(x)$ and $D_{1b\_max}(x)$, clipping to 0 and 1 when out of range. Next, in our experiments, if Adam is used, the discriminator's loss sometimes collapses. When it collapses, the cosine of the angle between the update direction and the gradient direction given by Adam becomes negative. It means that the discriminator's loss gradient is unstable, so it is not suitable to use a momentum-based optimization algorithm such as Adam. Since the RMSProp optimization algorithm is suitable for gradient instability, we change the adaptive optimization algorithm from Adam to RMSProp to make our model stable for amplitude feature maps from WGAN. In addition, we do not use the Wassertein distance as an optimization method to train the GAN model. The training process is shown in Algo. \ref{alg_afdcgan_training}.

\begin{algorithm}
	\caption{Training of AF-DCGAN Model.}\label{alg_afdcgan_training}
	\begin{algorithmic}[1]\label{algo}
		\Require $\Phi_i^{'}$ from initial fingerprint database $\Phi^{'}$, the learning rate $LR$, the clipping parameter $c$, the batch size $bs$, and the number of iterations of the discriminator per generator iteration $f_{d}$.
		
		\State Randomly initialize discriminator parameters $w_0$, generator's parameters $\theta _0$
		
		\ForAll {training iterations}
		\ForAll {$t$ in $f_{d}$}
		\State Sample mini-batch of $b_s$ examples $\{x^{(1)}, \cdots, x^{(b_s)}\}$ from the real data $\Phi _{i} ^{'}$
		\State Sample mini-batch of $b_s$ examples $\{z^{(1)}, \cdots, z^{(b_s)}\}$ from the noise prior $p_g(z)$
		\State $d_w\leftarrow \bigtriangledown _w[\frac{1}{bs}\sum_{i=1}^{bs}f_w(x^{(i)})-\frac{1}{bs}\sum_{i=1}^{bs}f_w(g_\theta(z^{(i)}))]$
		\State $w \leftarrow w +LR \times RMSProp(w,d_w)$
		\State $w \leftarrow clip(w,-c,c)$
		\EndFor
		\State Sample mini-batch of $bs$ examples $\{z^{(1)}, \cdots, z^{(b_s)}\}$ from the noise prior $p_g(z)$
		\State $g_\theta \leftarrow -\bigtriangledown _\theta\frac{1}{bs}\sum_{i=1}^{bs}f_w(g_\theta(z^{(i)}))$
		\State $\theta \leftarrow \theta - LR \times RMSProp(\theta,g_{\theta})$
		\EndFor
		
		\State In the generator $G$, output $l$ feature maps with $256\times 256$ resolution constitute a set of $\Psi_i$.
	\end{algorithmic}
\end{algorithm}

\begin{figure}[htb]
\centering
\includegraphics[width=3in]{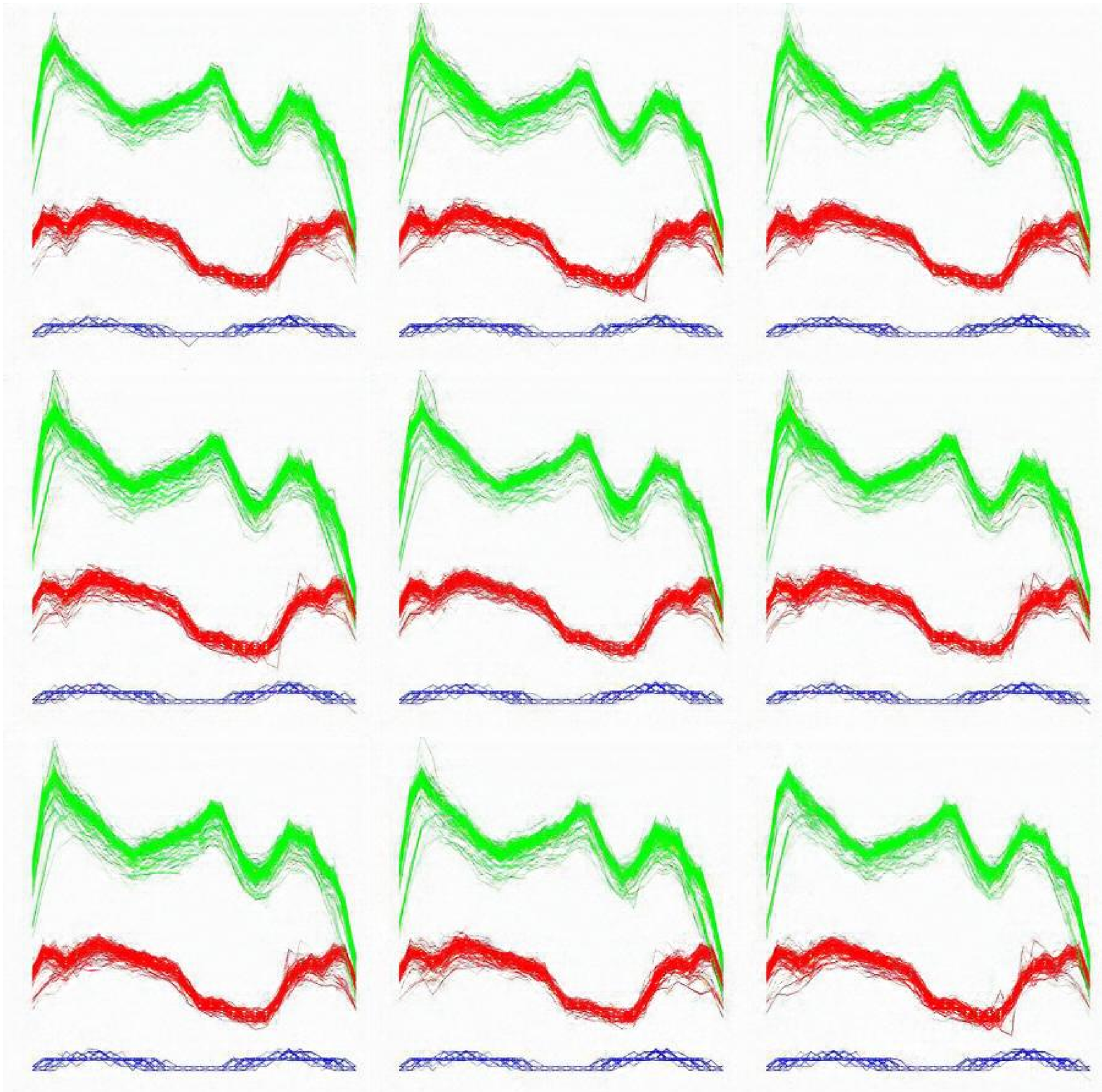}
\caption{CSI amplitude feature maps generated by the AF-DCGAN in the 60th epoch. It can be seen from the figure that there are obvious differences among all the subgraphs, which greatly improves the diversity of the samples.} \label{fig_afm_afdcgan_a}
\end{figure}

We enter $\Phi_i^{'}$ sequentially into the AF-DCGAN model to generate amplitude feature maps for all the locations to obtain the set of generated amplitude feature maps as follows:
\begin{equation}
	\Psi=\{\Psi_1,\cdots,\Psi_i,\cdots,\Psi_M\}
\end{equation}
where $\Psi_i$ denotes the amplitude feature maps of the $N_t\times N_r$ links generated by the generator corresponding to the $i$-th reference point. Example amplitude feature maps generated by the generator of a well-trained model are shown in Fig.\ref{fig_afm_afdcgan_a}. They look roughly similar, however, if we look at the details of generated feature maps, we can notice a lot of differences.

Thus far, we have used the AF-DCGAN to generate more amplitude feature maps, extending the initial fingerprint database and greatly reducing the manual labor. We eventually obtain the expanded fingerprint database $\Gamma=\{\Phi^{'},\Psi\}$.

\section{Validation Experiment} \label{experiment}
Experiments are conducted to evaluate the performance of the proposed fingerprint database extension method.

\subsection{Experiment Methodology}
The experiment is conducted in the classroom, as shown in Fig.~\ref{fig_experiment_environment}. The circumscribed rectangle of the classroom is taken as the indoor positioning area ($7m \times 7m$), which is evenly divided into $49$ square grids. The center point of each cell is used as a reference point to form a set of reference points. The distance between two adjacent reference points is one meter. This distance is selected for the following reasons. As shown in Fig.~\subref*{fig_csi_packets_a} and Fig.~\subref*{fig_csi_packets_b}, the two positions are one meter apart, and we can see the difference in their amplitude maps, although they are similar. When the distances are too close, the receiving paths of the adjacent locations are similar; consequently, the features of the amplitude feature maps in the fingerprint database will be highly similar, which will reduce the localization accuracy. As shown in Fig.~\subref*{fig_csi_packets_a} and Fig.~\subref*{fig_csi_packets_c}, the two positions are five meters apart, and we can see that their amplitude maps are significantly different. When the distances are too far apart, the feature differences of the amplitude feature maps of adjacent locations will become larger, making certain features difficult to match and resulting in reduced localization accuracy. 

In the classroom, we deployed a TL-WR742N wireless router as the transmitter, operating in IEEE 802.11n AP mode and equipped with one transmitting antenna, and a ThinkPad x201 laptop equipped with Intel Wireless Link 5300 NICs (IWL5300) as receivers, with three receive antennas. The Laptop’s operating system is Ubuntu 16.04, installed on a custom system kernel with a modified network driver. The firmware and driver of IWL5300 are modified to export the CSI of each packet's predicted IEEE 802.11 packet delivery from wireless channel measurements made by the ‘Linux 802.11n CSI Tool’ \cite{33} and containing information about all subcarriers. The router was placed at the front of the measurement area. The receiving antennas were placed behind the measurement area to cover the whole area. To collect the CSI data, \emph{Ping} commands were executed on the laptop every 0.4 s to generate network traffic \footnote{The captured CSI data can be downloaded at https://github.com/quheng54/dataset-for-AF-DCGAN}. The AF-DCGAN model was implemented on TensorFlow and accelerated by a GPU (GeForce GTX 1060).

\begin{figure}[htb]
\centering
\includegraphics[width=2.5in]{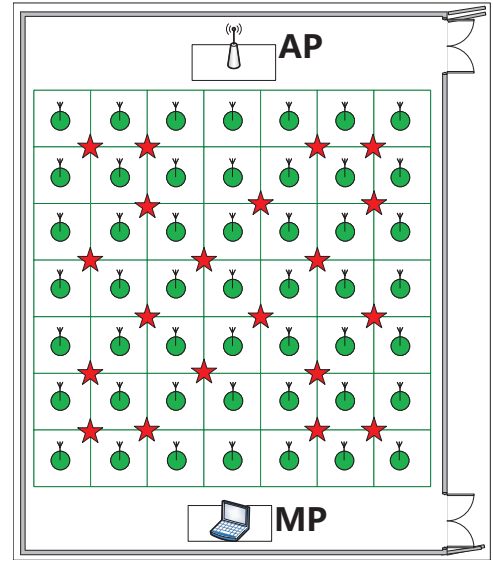}
\caption{Layout of the classroom for training/test positions. }
\label{fig_experiment_environment_layout}
\end{figure}

During the experiments, we collected CSI data first. Then, after processing it into amplitude feature maps, we use AF-DCGAN to generate the additional amplitude feature maps. We set the learning rate of the model $LR=0.0002$, cutting parameter $c=0.01$, batch size $bs=49$, and the number of iterations of discriminator per generator iteration $f_{d} = 2$.  When transmitting CSI data, the volunteers stood at the reference points (shown in Fig.~\ref{fig_experiment_environment_layout} as the green circles). At each reference point, 5000 data samples are collected. There are three links and each link contains 30 groups of subcarriers. Thus, each CSI sample has $1\times 3\times 30$ dimensions. For each reference point, we randomly select 100 samples from the $5000$ samples for $10000$ times to draw the amplitude feature maps. After pixel transformation, the obtained amplitude feature maps of the three links form the initial fingerprint database. Then, we use the AF-DCGAN model corresponding to each position to generate $10000$ additional amplitude maps for each reference point that reflect the same type of reference point to extend the initial database. In this way, the diversity of samples generated at a single location can be increased, and the localization accuracy can be improved. We call the extended fingerprint database consisting of the newly generated feature maps GOAFM.

\begin{figure}[htb]
	\centering
	\includegraphics[width=3.5in]{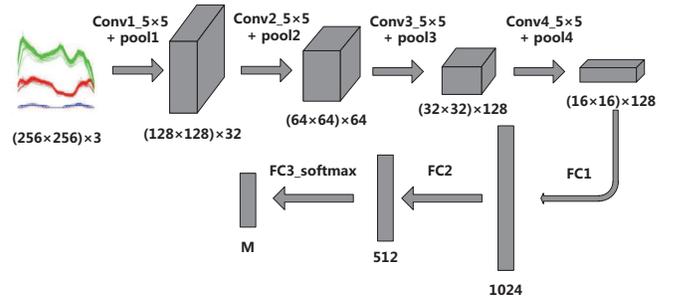}
	\caption{Architecture of AF-CNN model.}
	\label{fig_AF_cnn_model}
\end{figure}

During the localization step, a deep learning method is used to classify the amplitude feature maps, similar to device-free indoor localization in \cite{7765094} and Deepfi in \cite {35}, \cite{7438932}, which are commonly used for localization. For clarity, the architecture is illustrated in Fig.~\ref{fig_AF_cnn_model}, and is named Amplitude Feature CNN (AF-CNN).
We treat the AF-CNN as a classification model to determine the reference point to which a test point belongs. During the experiments, we select $20$ points in the experimental area as test points that differ from the reference point (shown as the red stars in Fig.~\ref{fig_experiment_environment_layout}). Each point has $5000$ samples. The test-sample point data are processed in the same manner as explained above to obtain the amplitude feature maps required for the test. Then, we put the amplitude feature maps into the trained classification model for testing. The test results output the first four best-matching reference points. The first four best-matching reference points are selected because the test points are often surrounded by four reference points; thus, using the center of the resulting first four highest matching reference points improves the localization accuracy. Finally, we calculate the geometric center of the first four highest matching points as the localization result.

Finally, we verify the superiority of the extension fingerprint database by comparing the localization results using the initial fingerprint database and the fingerprint database expanded by AF-DCGAN. We also compare the localization results using the initial fingerprint database and the fingerprint database expanded by adding noise (ADNOI). During the experiments, we simply add some Gaussian random noise to the sampled data in original amplitude feature maps, with , to generate an augmented fingerprint database. We name it 100\% ADNOI when the same amount of feature maps are supplemented, and 200\% ADNOI when adding twice the amount of maps.

Additionally, we compare our method to existing methods for building a fingerprint database with GPR and with the existing localization methods FIFS \cite{34} and DeepFi \cite{35} to demonstrate our method's competitiveness.

\subsection{Localization performance}
\begin{figure}[htb]
\centering
\subfloat[Initial database.]{
  \includegraphics[width=3in]{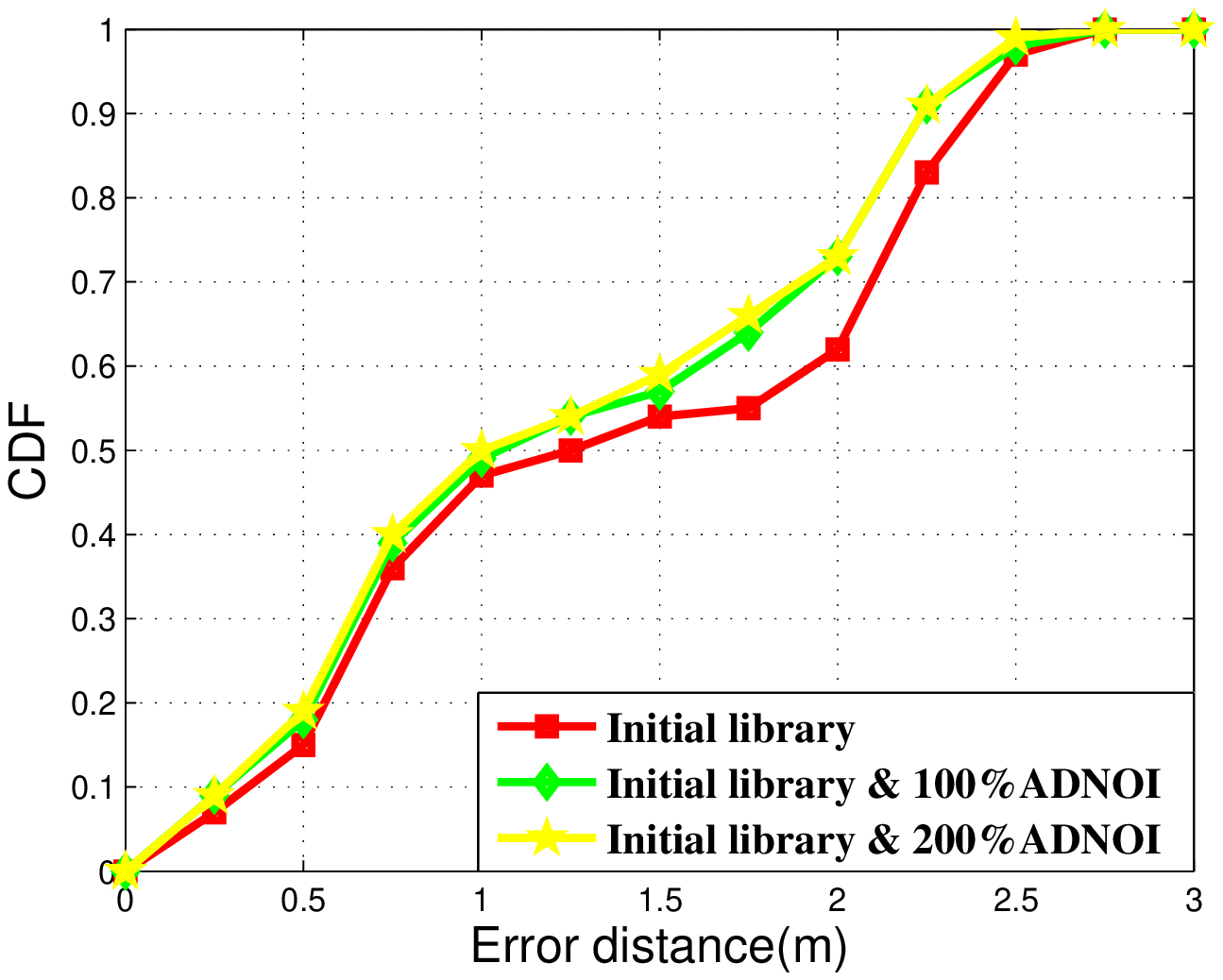}\label{fig_localization_cdf_initial}}\hfill
\subfloat[Initial database and GOAFM.]{
  \includegraphics[width=3in]{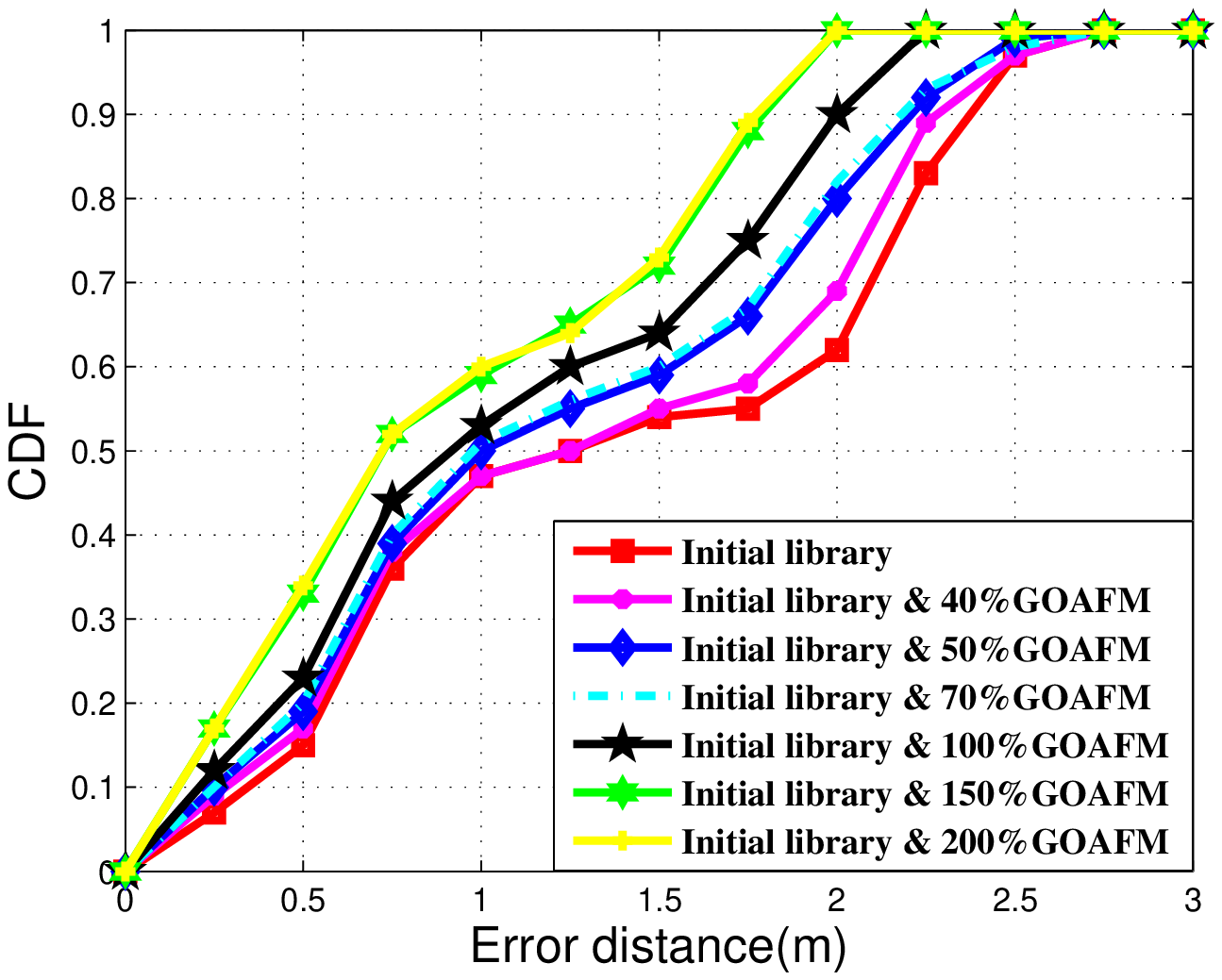}\label{fig_localization_cdf_goafm}}\hfill
\subfloat[ADNOI and GOAFM.]{
  \includegraphics[width=3in]{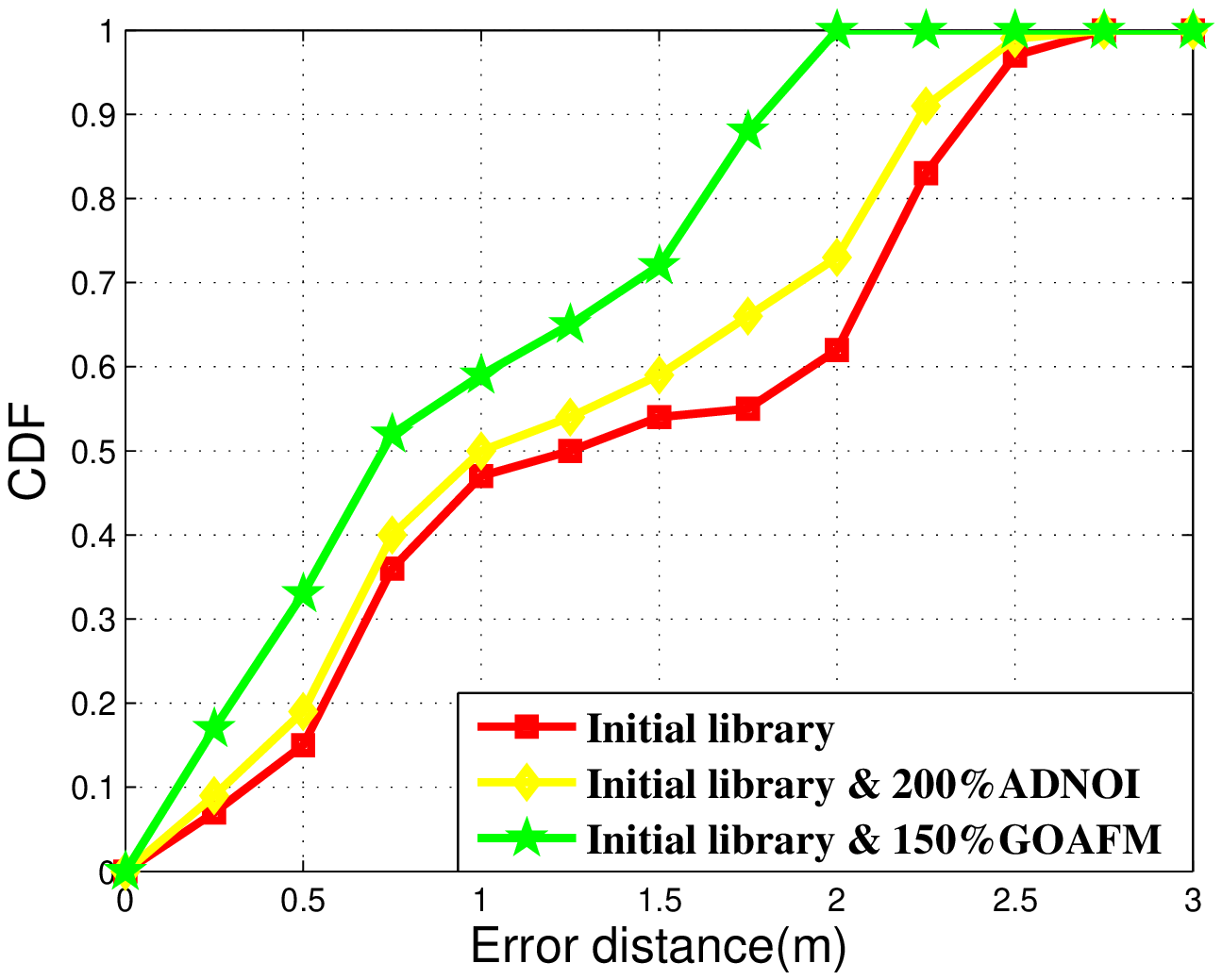}\label{fig_localization_cdf_initial_goafm}}\hfill
\caption{As the number of samples increases, the localization accuracy of the fingerprints database increase.} \label{fig_localization_cdf}
\end{figure}

(1) Performance on different localization methods: We use AF-CNN to conduct the localization experiment. Fig.~\subref*{fig_localization_cdf_initial} shows the cumulative distribution function (CDF) of the localization error for the initial fingerprint database and the extended fingerprint database by adding noise. The red line represents the CDF curve using the initial fingerprint database, which has an average error of $1.34$m. For this fingerprint database, the minimum error distance is $0.12$m, and the maximum error distance is $2.68$m. The probability of an error distance within $1$m is 47\%, within $2$m, 62\%, and within $3$m, 100\%. The green line represents the CDF curve using the initial database with an additional 100\% of ADNOI, and its average error is $1.25$m. In this condition, the minimum error distance is $0.05$m, and the maximum error distance is $2.47$m. The probability of an error distance within $1$m is 49\%, within $2$m, 73\%, and within $3$m 100\%. The yellow line represents the CDF curve when the test points are located using a fingerprint database with an added 200\% of fingerprints of amplitude feature maps generated by adding noise for each point, and its average error is $1.23$m. For this fingerprint database, the minimum error distance is $0.06$m, and the maximum error distance is $2.45$m. The probability of an error distance within $1$m is 50\%, within $2$m, 73\%, and within $3$m 100\%.

These results demonstrate the validity of the database generated using ADNOI. We can use ADNOI to augment the database without expending human effort while also improving the localization accuracy. With 200\% ADNOI is added, the sample size of the fingerprint database is saturated and the error is no longer reduced.

As shown in Fig.~\subref*{fig_localization_cdf_goafm}, the red line represents the CDF curve using the initial fingerprint database. The blue line represents the CDF curve using the initial database with an additional 50\% of GOAFM, and its average error is $1.21$m. In this condition, the minimum error distance is $0.06$m, and the maximum error distance is $2.63$m. The probability of an error distance within $1$m is 50\%, within $2$m, 80\%, and within $3$m 100\%. The black line represents the CDF curve when the test points are located using a fingerprint database with an added 100\% of fingerprints of amplitude feature maps generated by AF-DCGAN for each point, and its average error is $1.18$m. For this fingerprint database, the minimum error distance is $0.04$m, and the maximum error distance is $2.23$m. The probability of an error distance within $1$m is 53\%, within $2$m, 90\%, and within $3$m 100\%. The green line represents the CDF curve with 150\% of generated fingerprints of amplitude feature maps added to the initial database, and its average error is $0.92$m. In this condition, the minimum error distance is $0.03$m, and the maximum error distance is $1.98$m. The probability of an error distance within $1$m is 59\%, within $2$m, 100\%, and within $3$m, 100\%. The yellow line represents the CDF curve with 200\% of generated fingerprints of amplitude feature maps added to the initial database, it has an average error of $0.92$m. For this fingerprint database, the minimum error distance is $0.03$m, and the maximum error distance is $1.96$m. The probability of an error distance within $1$m is 59\%, within $2$m, 100\%, and within $3$m, 100\%. The location results are shown in Tables \ref{table_localization_error} and \ref{table_localization_cdf}. Error is represented by the mean error distance. As shown in Fig.~\subref*{fig_localization_cdf_goafm}, for the initial fingerprint database, having more fingerprint database samples available improves the positioning accuracy. After adding the generated amplitude feature maps to the initial database, the localization accuracy improve. After adding 50\% GOAFM to the initial database, the accuracy is $1.21$m, an improvement of 9.70\%. After adding 100\% GOAFM to the initial database, the accuracy is $1.18$m, an improvement of 16.00\%, and after adding 150\% GOAFM to the initial database, the accuracy is $0.92$m, an improvement of 31.30\%. After 150\% GOAFM is added, the sample size of the fingerprint database is basically saturated, and the error is no longer reduced. So, after adding 200\% GOAFM to the initial database, the accuracy is the same as the 150\% GOAFM added to the initial database.

These results demonstrate the validity of the database generated using AF-DCGAN. In other words, we can use AF-DCGAN to augment the database without expending human effort while also improving the localization accuracy. As shown in Fig.~\subref*{fig_localization_cdf_initial_goafm}, adding GOAFM generated by the AF-DCGAN is better than adding ADNOI to the initial database.

\begin{figure}[tb]
\centering
\subfloat[Initial database with FIFS and DeepFi.]{
  \includegraphics[width=3in]{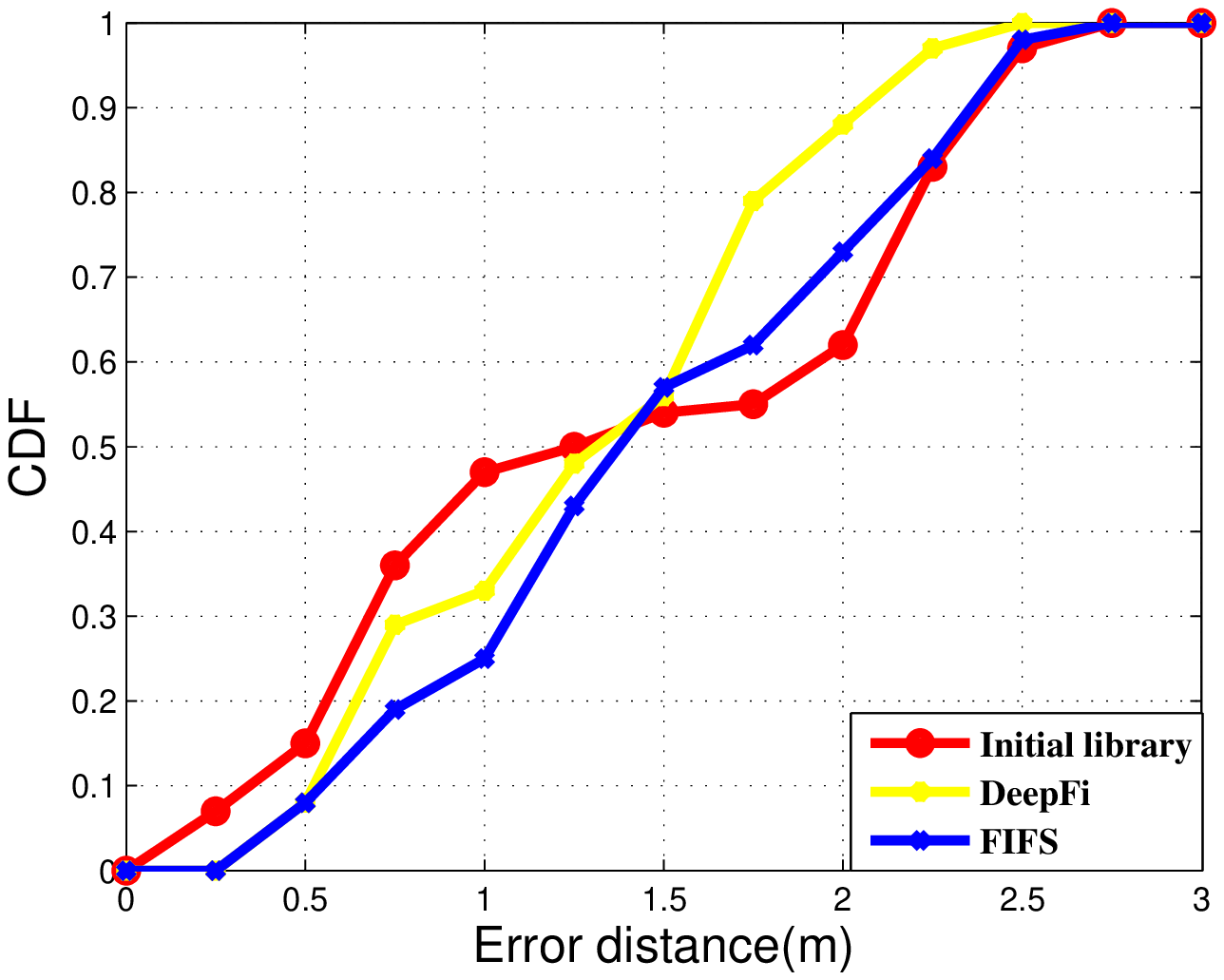}\label{fig_localization_performance_inital}}\hfill
\subfloat[Initial database and GOAFM with FIFS and DeepFi.]{
  \includegraphics[width=3in]{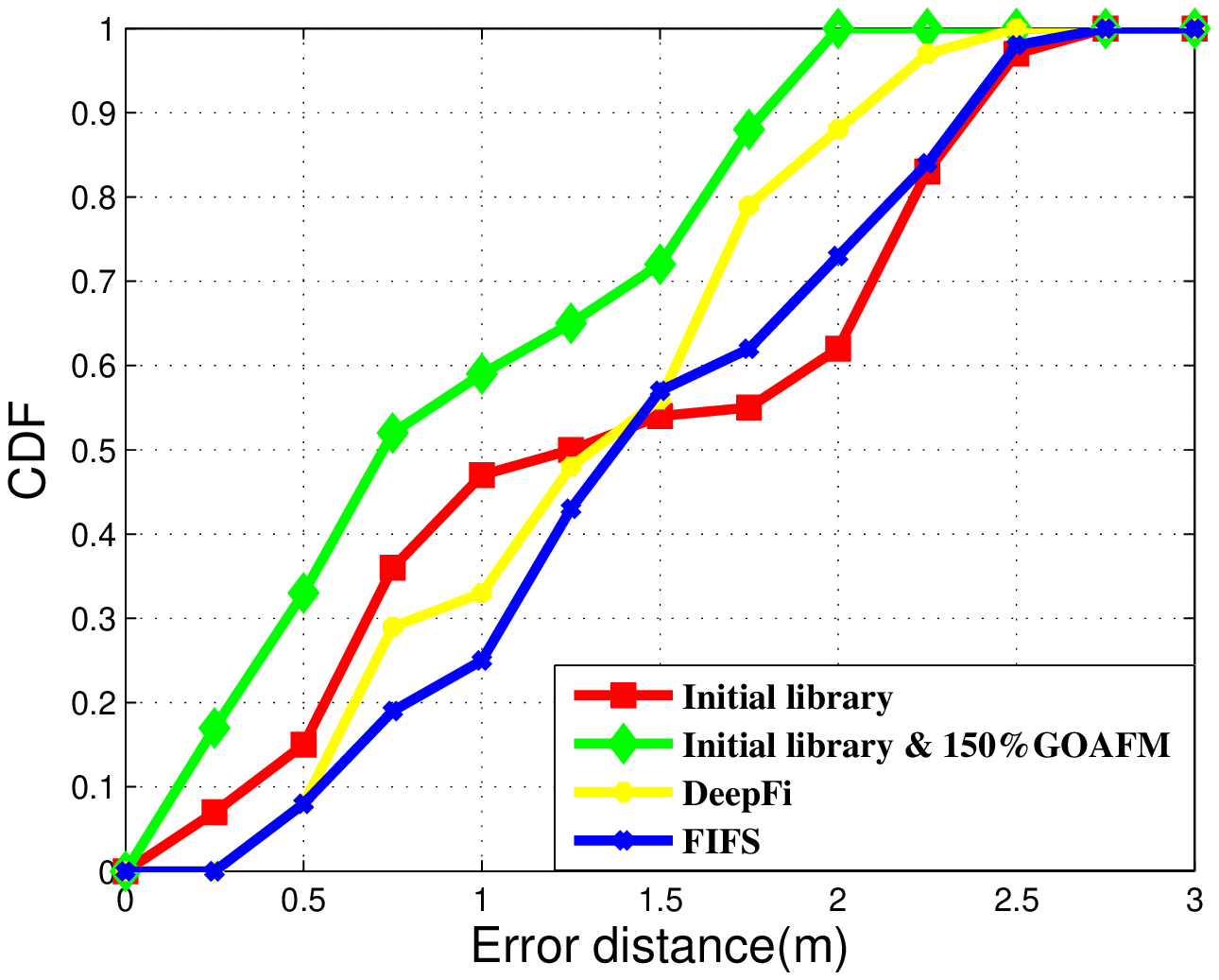}\label{fig_localization_performance_inital_GOAFM}}\hfill
\caption{Comparison to other localization methods.} \label{fig_database_construction_performance}
\end{figure}

We also evaluate the performance of our localization method based on our fingerprint database by comparing it with the existing localization methods FIFS and DeepFi. We conduct these three experiments in the classroom environment as illustrated in Fig.~\ref{fig_experiment_environment}. As shown in Fig.~\subref*{fig_localization_performance_inital}, Fig.~\subref*{fig_localization_performance_inital_GOAFM}, Tables \ref{table_localization_error} and \ref{table_localization_cdf}, our method is superior to FIFS and DeepFi. When we used the initial database for positioning, compared to FIFS, the localization accuracy improved by 3.60\%. When we added all the GOAFM to the initial database, the localization accuracy improved by 15.11\%. When we added 150\% of the GOAFM to the initial database, the localization accuracy improved by 33.81\%. In other words, we are able to significantly improve the localization accuracy without expending additional labor to collect more CSI data. As shown in Fig.~\subref*{fig_localization_performance_inital} and Fig.~\subref*{fig_localization_performance_inital_GOAFM}, the localization accuracy of the initial database is not better compared with DeepFi, but after adding 150\% GOAFM, its positioning accuracy increases by 24.59\%, even without consuming additional labor to collect more CSI data.

	\begin{table}[htbp]
		\centering
		\caption{Localization Error of Different Methods}\label{table_localization_error}
		\centering
		\begin{tabular}{c|c|c|c}
        \hline
		Method 							& Min. 	& Max. 	& Mean \\\hline
		Initial library  				& 0.12m & 2.68m & 1.34m \\
		Initial library and 100\% ADNOI	& 0.05m & 2.47m & 1.25m \\
		Initial library and 200\% ADNOI & 0.06m & 2.45m & 1.23m \\\hline
		Initial library and 50\% GOAFM	& 0.06m & 2.63m & 1.21m \\
		Initial library and 100\% GOAFM	& 0.04m & 2.23m & 1.18m \\
		Initial library and 150\% GOAFM & 0.03m & 1.98m & 0.92m \\
		Initial library and 200\% GOAFM & 0.03m & 1.96m & 0.92m \\\hline
		FIFS							& 0.27m	& 2.71m	& 1.39m \\\hline
		DeepFI							& 0.29m	& 2.47m	& 1.22m \\\hline
		\end{tabular}
	\end{table}

	\begin{table}[htbp]
		\centering
		\caption{Localization Range Probability of Different Methods}\label{table_localization_cdf}
		\centering
		\begin{tabular}{c|c|c|c}
        \hline
		Method 							& 1m 	& 2m 	& 3m \\\hline
		Initial library  				& 47\% 	& 62\% 	& 100\% \\
		Initial library and 100\% ADNOI	& 49\% 	& 73\% 	& 100\% \\
		Initial library and 200\% ADNOI & 50\% 	& 73\%  & 100\% \\\hline
		Initial library and 50\% GOAFM	& 50\% 	& 80\% 	& 100\% \\
		Initial library and 100\% GOAFM	& 53\% 	& 90\% 	& 100\% \\
		Initial library and 150\% GOAFM & 59\% 	& 100\% & 100\% \\
		Initial library and 200\% GOAFM & 59\% 	& 100\% & 100\% \\\hline
		FIFS							& 23\% 	& 73\% 	& 100\% \\\hline
		DeepFI							& 33\% 	& 88\% 	& 100\% \\\hline
		\end{tabular}
	\end{table}

\begin{figure}[htb]
\centering
\subfloat[Initial database.]{
  \includegraphics[width=3in]{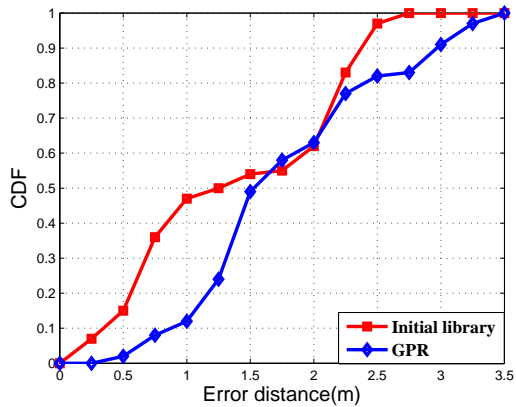}\label{fig_database_construction_performance_intial_GPR}}\hfill
\subfloat[GOAFM.]{
  \includegraphics[width=3in]{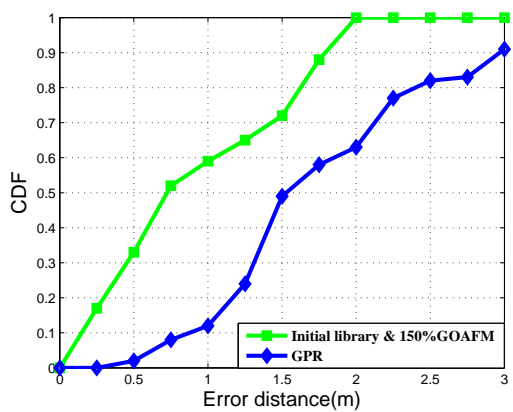}\label{fig_database_construction_performance_intial_GOAFM_GPR}}\hfill
\caption{Comparison with the other database construction methods.} \label{fig_database_construction_performance}
\end{figure}

(2) Performance on different database construction methods: In this subsection, we evaluate the performance of our AF-DCGAN-based fingerprint database  by comparing it with existing methods of fingerprint database construction based on GPR. We conduct these two experiments in the research classroom using the same CSI dataset. For GPR modeling, we select 4 of the 30 carrier models that are more accurate. Then, we determine the geometric center of these 4 model positioning results as the final positioning result. The results shown in Fig.~\subref*{fig_database_construction_performance_intial_GPR}, Fig.~\subref*{fig_database_construction_performance_intial_GOAFM_GPR}, Tables \ref{table_localization_contruction_error} and \ref{table_localization_construction_cdf} reveal that our fingerprint database construction method is superior to GPR. When we use the initial database for positioning, compared to GPR , the localization accuracy improved by 23.42\%. When we add the 150\% GOAFM to the initial database, the localization accuracy improved by 47.43\%. Moreover, the localization accuracy is significantly improved without expending more labor to collect the CSI data because of the AF-DCGAN.

	\begin{table}[htbp]
		\centering
		\caption{Localization Error of Different Database Construction Methods}\label{table_localization_contruction_error}
		\centering
		\begin{tabular}{c|c|c|c}
        \hline
		Method 							& Min. 	& Max. 	& Mean \\\hline
		Initial library and 50\% GOAFM	& 0.06m & 2.63m & 1.21m \\
		Initial library and 100\% GOAFM	& 0.04m & 2.23m & 1.18m \\
		Initial library and 150\% GOAFM	& 0.03m & 1.98m & 0.92m 
		\\\hline
		GPR								& 0.34m	& 3.34m	& 1.75m \\\hline
		\end{tabular}
	\end{table}

	\begin{table}[!t]
	\renewcommand{\arraystretch}{1.3}
		\caption{Localization Range Probability of Different Database Construction Methods}\label{table_localization_construction_cdf}
		\centering
		\begin{tabular}{c|c|c|c}
        \hline
		Method 							& 1m 	& 2m 	& 3m \\\hline
		Initial library and 50\% GOAFM	& 50\% 	& 80\% 	& 100\% \\
		Initial library and 100\% GOAFM	& 53\% 	& 90\% 	& 100\% \\
		Initial library and 150\% GOAFM & 59\% 	& 100\% 	& 100\% \\\hline
		GPR								& 24\% 	& 63\% 	& 83\% \\\hline
		\end{tabular}
	\end{table}

\section{CONCLUSION} \label{conclusion}

In this paper, a novel approach was proposed to reduce the collection of WiFi fingerprints. The new AF-DCGAN model based on GAN was used to generate additional amplitude feature maps similar to those in the reference database, which effectively increased the number of samples in the training set. We conducted exhaustive tests to demonstrate the performance of the proposed method, and the results showed the superiority of AF-DCGAN over the existing methods of building fingerprint database and localization methods, i.e. the accuracy of indoor WiFi positioning was improved without increasing the labor involved in building the fingerprint databases. The proposed scheme has limitations to be addressed in the future. For example, CSI data is highly susceptible to environmental changes, and localization accuracy will decrease when the indoor environment changes significantly.

We envision localization to play a vital role in the future society by its ability to offer enriched applications. To make the localization more accurate we will consider more types of information such as phase information in CSI data besides the amplitude of CSI data in the future. Pretrained networks and more complex CNN will be investigated for further performance improvement.  Besides, the GAN-based scheme will be compared with other popular data generation method such as SMOTE \cite{articleSMOTE} in terms of accuracy and computational complexity.

%

\section*{Acknowledgment}
The corresponding author of this manuscript is Jie Li. This research is supported in part by National Natural Science Foundation of China, Grant No. 51877060, ANHUI Province Key Laboratory of Affective Computing \& Advanced Intelligent Machine, Grant No.ACAIM180102, and the Fundamental Research Funds for the Central Universities, Grant No. JZ2018HGTB0253, JZ2019HGTB0089 and PA2019GDQT0006, and State Grid Science and Technology Project (Research and application of key Technologies for integrated substation intelligent operation and maintenance based on the fusion of heterogeneous network and heterogeneous data).
\bibliographystyle{IEEEtran}
\bibliography{GAN}

\begin{thebibliography}{10}
\providecommand{\url}[1]{#1}
\csname url@samestyle\endcsname
\providecommand{\newblock}{\relax}
\providecommand{\bibinfo}[2]{#2}
\providecommand{\BIBentrySTDinterwordspacing}{\spaceskip=0pt\relax}
\providecommand{\BIBentryALTinterwordstretchfactor}{4}
\providecommand{\BIBentryALTinterwordspacing}{\spaceskip=\fontdimen2\font plus
\BIBentryALTinterwordstretchfactor\fontdimen3\font minus
  \fontdimen4\font\relax}
\providecommand{\BIBforeignlanguage}[2]{{%
\expandafter\ifx\csname l@#1\endcsname\relax
\typeout{** WARNING: IEEEtran.bst: No hyphenation pattern has been}%
\typeout{** loaded for the language `#1'. Using the pattern for}%
\typeout{** the default language instead.}%
\else
\language=\csname l@#1\endcsname
\fi
#2}}
\providecommand{\BIBdecl}{\relax}
\BIBdecl

\bibitem{3}
Z.~Gu, Z.~Chen, Y.~Zhang, Y.~Zhu, M.~L, and A.~Chen, ``Reducing fingerprint
  collection for indoor localization,'' \emph{Computer Communications},
  vol.~83, 09 2015.

\bibitem{8353839}
X.~Guo, L.~Li, N.~Ansari, and B.~Liao, ``Accurate wifi localization by fusing a
  group of fingerprints via global fusion profile,'' \emph{IEEE Transactions on
  Vehicular Technology}, pp. 1--1, 2018.

\bibitem{8326317}
V.~Singh, G.~Aggarwal, and B.~V.~S. Ujwal, ``Ensemble based real-time indoor
  localization using stray wifi signal,'' in \emph{2018 IEEE International
  Conference on Consumer Electronics (ICCE)}, Jan 2018, pp. 1--5.

\bibitem{4}
Q.~Li, B.~Chu, Z.~Wu, W.~Sun, L.~Chen, J.~Li, and Z.~Liu, ``Rmds: Ranging and
  multidimensional scalingâ“based anchor-free localization in large-scale
  wireless sensor networks with coverage holes,'' \emph{International Journal
  of Distributed Sensor Networks}, vol.~13, pp. 1--12, 08 2017.

\bibitem{5}
M.~Youssef and A.~Agrawala, ``The horus wlan location determination system,''
  in \emph{Proceedings of the 3rd International Conference on Mobile Systems,
  Applications, and Services}, ser. MobiSys '05.\hskip 1em plus 0.5em minus
  0.4em\relax New York, NY, USA: ACM, 2005, pp. 205--218.

\bibitem{6}
W.~Sun, X.~Yuan, J.~Wang, Q.~Li, L.~Chen, and D.~Mu, ``End-to-end data delivery
  reliability model for estimating and optimizing the link quality of
  industrial wsns,'' \emph{IEEE Transactions on Automation Science and
  Engineering}, vol.~PP, no.~99, pp. 1--11, 2017.

\bibitem{8304587}
S.~Shi, S.~Sigg, L.~Chen, and Y.~Ji, ``Accurate location tracking from
  csi-based passive device-free probabilistic fingerprinting,'' \emph{IEEE
  Transactions on Vehicular Technology}, vol.~67, no.~6, pp. 5217--5230, June
  2018.

\bibitem{7}
T.~Pulkkinen, T.~Roos, and P.~Myllym\"{a}ki, ``Semi-supervised learning for
  wlan positioning,'' in \emph{Proceedings of the 21th International Conference
  on Artificial Neural Networks - Volume Part I}, ser. ICANN'11.\hskip 1em plus
  0.5em minus 0.4em\relax Berlin, Heidelberg: Springer-Verlag, 2011, pp.
  355--362.

\bibitem{8}
S.~Liu, H.~Luo, and S.~Zou, ``A low-cost and accurate indoor localization
  algorithm using label propagation based semi-supervised learning,'' in
  \emph{2009 Fifth International Conference on Mobile Ad-hoc and Sensor
  Networks}, Dec 2009, pp. 108--111.

\bibitem{goodfellow2014generative}
I.~Goodfellow, J.~Pouget-Abadie, M.~Mirza, B.~Xu, D.~Warde-Farley, S.~Ozair,
  A.~Courville, and Y.~Bengio, ``Generative adversarial nets,'' in
  \emph{Advances in neural information processing systems}, 2014, pp.
  2672--2680.

\bibitem{cao2019recent}
Y.-J. Cao, L.-L. Jia, Y.-X. Chen, N.~Lin, C.~Yang, B.~Zhang, Z.~Liu, X.-X. Li,
  and H.-H. Dai, ``Recent advances of generative adversarial networks in
  computer vision,'' \emph{IEEE Access}, vol.~7, pp. 14\,985--15\,006, 2019.

\bibitem{9}
S.~He, S.~H.~G. Chan, L.~Yu, and N.~Liu, ``Fusing noisy fingerprints with
  distance bounds for indoor localization,'' in \emph{2015 IEEE Conference on
  Computer Communications (INFOCOM)}, April 2015, pp. 2506--2514.

\bibitem{8307353}
X.~Sun, X.~Gao, G.~Y. Li, and W.~Han, ``Single-site localization based on a new
  type of fingerprint for massive mimo-ofdm systems,'' \emph{IEEE Transactions
  on Vehicular Technology}, vol.~67, no.~7, pp. 6134--6145, July 2018.

\bibitem{11}
Z.~Xiao, H.~Wen, A.~Markham, and N.~Trigoni, ``Lightweight map matching for
  indoor localisation using conditional random fields,'' in \emph{Proceedings
  of the 13th International Symposium on Information Processing in Sensor
  Networks}, ser. IPSN '14.\hskip 1em plus 0.5em minus 0.4em\relax Piscataway,
  NJ, USA: IEEE Press, 2014, pp. 131--142.

\bibitem{12}
P.~Zhang, Q.~Zhao, Y.~Li, X.~Niu, Y.~Zhuang, and J.~Liu, ``Collaborative wifi
  fingerprinting using sensor-based navigation on smartphones,''
  \emph{Sensors}, vol. 2015, pp. 17\,534--17\,557, 07 2015.

\bibitem{7857072}
C.~K. Sung, F.~de~Hoog, Z.~Chen, P.~Cheng, and D.~C. Popescu, ``Interference
  mitigation based on bayesian compressive sensing for wireless localization
  systems in unlicensed band,'' \emph{IEEE Transactions on Vehicular
  Technology}, vol.~66, no.~8, pp. 7038--7049, Aug 2017.

\bibitem{16}
J.~Jun, L.~He, Y.~Gu, W.~Jiang, G.~Kushwaha, V.~A, L.~Cheng, C.~Liu, and
  T.~Zhu, ``Low-overhead wifi fingerprinting,'' \emph{IEEE Transactions on
  Mobile Computing}, vol.~PP, no.~99, pp. 1--1, 2017.

\bibitem{17}
Y.~Shu, Y.~Huang, J.~Zhang, P.~Coué, P.~Cheng, J.~Chen, and K.~G. Shin,
  ``Gradient-based fingerprinting for indoor localization and tracking,''
  \emph{IEEE Transactions on Industrial Electronics}, vol.~63, no.~4, pp.
  2424--2433, April 2016.

\bibitem{18}
J.~Niu, B.~Wang, L.~Shu, T.~Q. Duong, and Y.~Chen, ``Zil: An energy-efficient
  indoor localization system using zigbee radio to detect wifi fingerprints,''
  \emph{IEEE Journal on Selected Areas in Communications}, vol.~33, no.~7, pp.
  1431--1442, July 2015.

\bibitem{19}
S.~Yoon, K.~Lee, Y.~Yun, and I.~Rhee, ``Acmi: Fm-based indoor localization via
  autonomous fingerprinting,'' \emph{IEEE Transactions on Mobile Computing},
  vol.~15, no.~6, pp. 1318--1332, June 2016.

\bibitem{7980032}
L.~Chang, J.~Xiong, Y.~Wang, X.~Chen, J.~Hu, and D.~Fang, ``iupdater: Low cost
  rss fingerprints updating for device-free localization,'' in \emph{2017 IEEE
  37th International Conference on Distributed Computing Systems (ICDCS)}, June
  2017, pp. 900--910.

\bibitem{21}
D.~Milioris, M.~Bradonjić, and P.~Mühlethaler, ``Building complete training
  maps for indoor location estimation,'' in \emph{2015 IEEE Conference on
  Computer Communications Workshops (INFOCOM WKSHPS)}, April 2015, pp. 75--76.

\bibitem{8003484}
C.~Wu, Z.~Yang, and C.~Xiao, ``Automatic radio map adaptation for indoor
  localization using smartphones,'' \emph{IEEE Transactions on Mobile
  Computing}, vol.~17, no.~3, pp. 517--528, March 2018.

\bibitem{23}
F.~Lemic, V.~Handziski, G.~Caso, L.~D. Nardis, and A.~Wolisz, ``Enriched
  training database for improving the wifi rssi-based indoor fingerprinting
  performance,'' in \emph{2016 13th IEEE Annual Consumer Communications
  Networking Conference (CCNC)}, Jan 2016, pp. 875--881.

\bibitem{24}
C.~He, S.~Guo, Y.~Wu, and Y.~Yang, ``A novel radio map construction method to
  reduce collection effort for indoor localization,'' \emph{Measurement},
  vol.~94, 08 2016.

\bibitem{25}
Y.~Cho, J.~Kim, M.~Ji, Y.~Lee, and S.~Park, ``Gpr based wi-fi radio map
  construction from real/virtual indoor dynamic surveying data,'' in \emph{2013
  13th International Conference on Control, Automation and Systems (ICCAS
  2013)}, Oct 2013, pp. 712--714.

\bibitem{26}
S.~Kumar, R.~M. Hegde, and N.~Trigoni, ``Gaussian process regression for
  fingerprinting based localization,'' \emph{Ad Hoc Networks}, vol.~51, pp. 1
  -- 10, 2016.

\bibitem{27}
H.~Hu, W.~Zhou, Z.~Wen, Y.~Sun, and B.~Yin, ``Efficient radio map construction
  based on low-rank approximation for indoor positioning,'' \emph{Mathematical
  Problems in Engineering}, vol. 2013, no.~1, pp. 1--9, 2013.

\bibitem{28}
S.~Ezpeleta, J.~M. Claver, J.~J. Perez-Solano, and J.~V. MartÃ­, ``Rf-based
  location using interpolation functions to reduce fingerprint mapping,''
  vol.~15, p. 27322, 10 2015.

\bibitem{Shrivastava_2017_CVPR}
S.~Ashish, P.~Tomas, T.~Oncel, S.~Joshua, W.~Wenda, and W.~Russell, ``Learning
  from simulated and unsupervised images through adversarial training,'' in
  \emph{The IEEE Conference on Computer Vision and Pattern Recognition (CVPR)},
  July 2017.

\bibitem{29}
R.~Zhou, X.~Lu, P.~Zhao, and J.~Chen, ``Device-free presence detection and
  localization with svm and csi fingerprinting,'' \emph{IEEE Sensors Journal},
  vol.~17, no.~23, pp. 7990--7999, Dec 2017.

\bibitem{Xie2015}
Y.~Xie, Z.~Li, and M.~Li, ``Precise power delay profiling with commodity
  wifi,'' in \emph{Proceedings of the 21st Annual International Conference on
  Mobile Computing and Networking}, ser. MobiCom '15.\hskip 1em plus 0.5em
  minus 0.4em\relax New York, NY, USA: ACM, 2015, pp. 53--64.

\bibitem{XIAO201773}
Y.~Xiao, S.~Zhang, J.~Cao, H.~Wang, and J.~Wang, ``Exploiting distribution of
  channel state information for accurate wireless indoor localization,''
  \emph{Computer Communications}, vol. 114, pp. 73 -- 83, 2017.

\bibitem{30}
l.~Goodfellow, J.~Pouget-Abadie, M.~Mirza, B.~B.~Xu, D.~Warde-Farley, S.~Ozair,
  A.~Courville, and Y.~Bengio, ``Generative adversarial nets,'' Z.~Ghahramani,
  M.~Welling, C.~Cortes, N.~D. Lawrence, and K.~Q. Weinberger, Eds.\hskip 1em
  plus 0.5em minus 0.4em\relax Curran Associates, Inc., 2014, pp. 2672--2680.

\bibitem{31}
A.~Radford, L.~Metz, and S.~Chintala, ``Unsupervised representation learning
  with deep convolutional generative adversarial networks,'' 11 2015.

\bibitem{32}
\BIBentryALTinterwordspacing
M.~Arjovsky, S.~Chintala, and L.~Bottou, ``Wasserstein {GAN},'' \emph{CoRR},
  vol. abs/1701.07875, 2017. [Online]. Available:
  \url{http://arxiv.org/abs/1701.07875}
\BIBentrySTDinterwordspacing

\bibitem{33}
D.~Halperin, W.~Hu, A.~Sheth, and D.~Wetherall, ``Tool release: Gathering
  802.11n traces with channel state information,'' \emph{ACM SIGCOMM CCR},
  vol.~41, no.~1, p.~53, Jan. 2011.

\bibitem{7765094}
J.~Wang, X.~Zhang, Q.~Gao, H.~Yue, and H.~Wang, ``Device-free wireless
  localization and activity recognition: A deep learning approach,'' \emph{IEEE
  Transactions on Vehicular Technology}, vol.~66, no.~7, pp. 6258--6267, July
  2017.

\bibitem{35}
X.~Wang, L.~Gao, S.~Mao, and S.~Pandey, ``Deepfi: Deep learning for indoor
  fingerprinting using channel state information,'' in \emph{2015 IEEE Wireless
  Communications and Networking Conference (WCNC)}, March 2015, pp. 1666--1671.

\bibitem{7438932}
X.~Wang, L.~Gao, and S.~Mao, ``Csi-based fingerprinting for indoor
  localization: A deep learning approach,'' \emph{IEEE Transactions on
  Vehicular Technology}, vol.~66, no.~1, pp. 763--776, Jan 2017.

\bibitem{34}
J.~Xiao, K.~Wu, Y.~Yi, and L.~M. Ni, ``Fifs: Fine-grained indoor fingerprinting
  system,'' in \emph{2012 21st International Conference on Computer
  Communications and Networks (ICCCN)}, July 2012, pp. 1--7.

\bibitem{articleSMOTE}
C.~Nitesh, B.~Kevin, O.~Lawrence, and P.~Kegelmeyer, ``Smote: Synthetic
  minority over-sampling technique,'' \emph{J. Artif. Intell. Res. (JAIR)},
  vol.~16, pp. 321--357, 01 2002.

\end{thebibliography}

\ifCLASSOPTIONcaptionsoff
  \newpage
\fi

\begin{IEEEbiography}[{\includegraphics[width=1in,height=1.25in,clip,keepaspectratio]{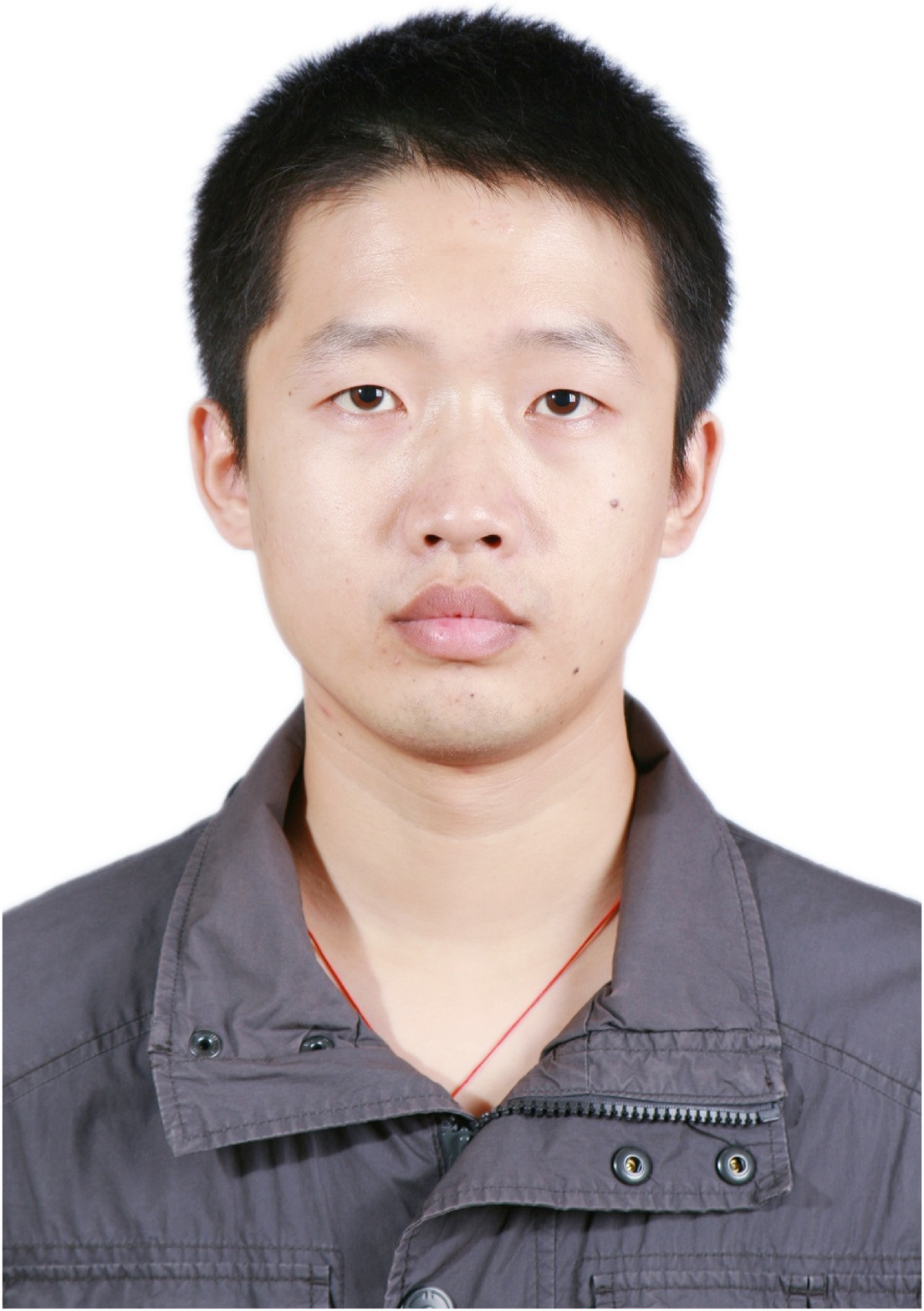}}]{Qiyue LI}
	(M'13) received the B.E., in electronic engineering from the Wuhan University, China, in 2003 and Ph.D. degree in communication and information system in University of Science and Technology, China. He is currently an Associate Professor at Hefei University of Technology, Anhui, China. From July 2008 to June 2011, he did postdoc research in school of computer science and technology, University of Science and Technology. His research interest includes wireless sensor networks and indoor localization using wireless networks. He is a member of IEEE and IEICE.
\end{IEEEbiography}

\begin{IEEEbiography}[{\includegraphics[width=1in,height=1.25in,clip,keepaspectratio]{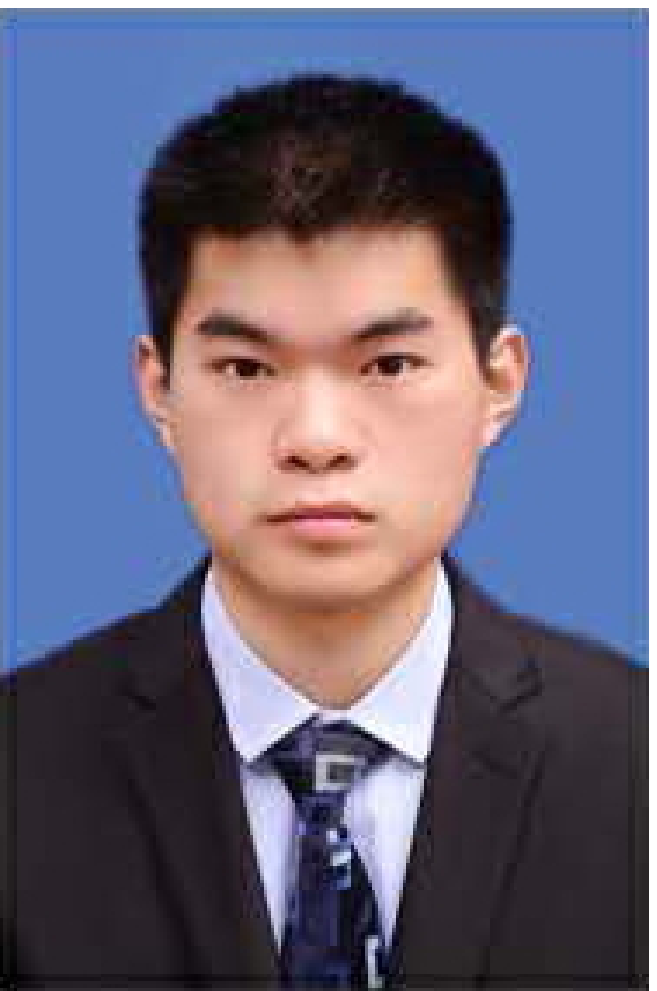}}]{Heng QU}
	received the B.E., in automation from the Hefei University of Technology in 2016. He is currently a postgraduate at Hefei University of Technology, Anhui, China since July 2016. His research interest includes deep learning and indoor localization. 
\end{IEEEbiography}

\begin{IEEEbiography}[{\includegraphics[width=1in,height=1.25in,clip,keepaspectratio]{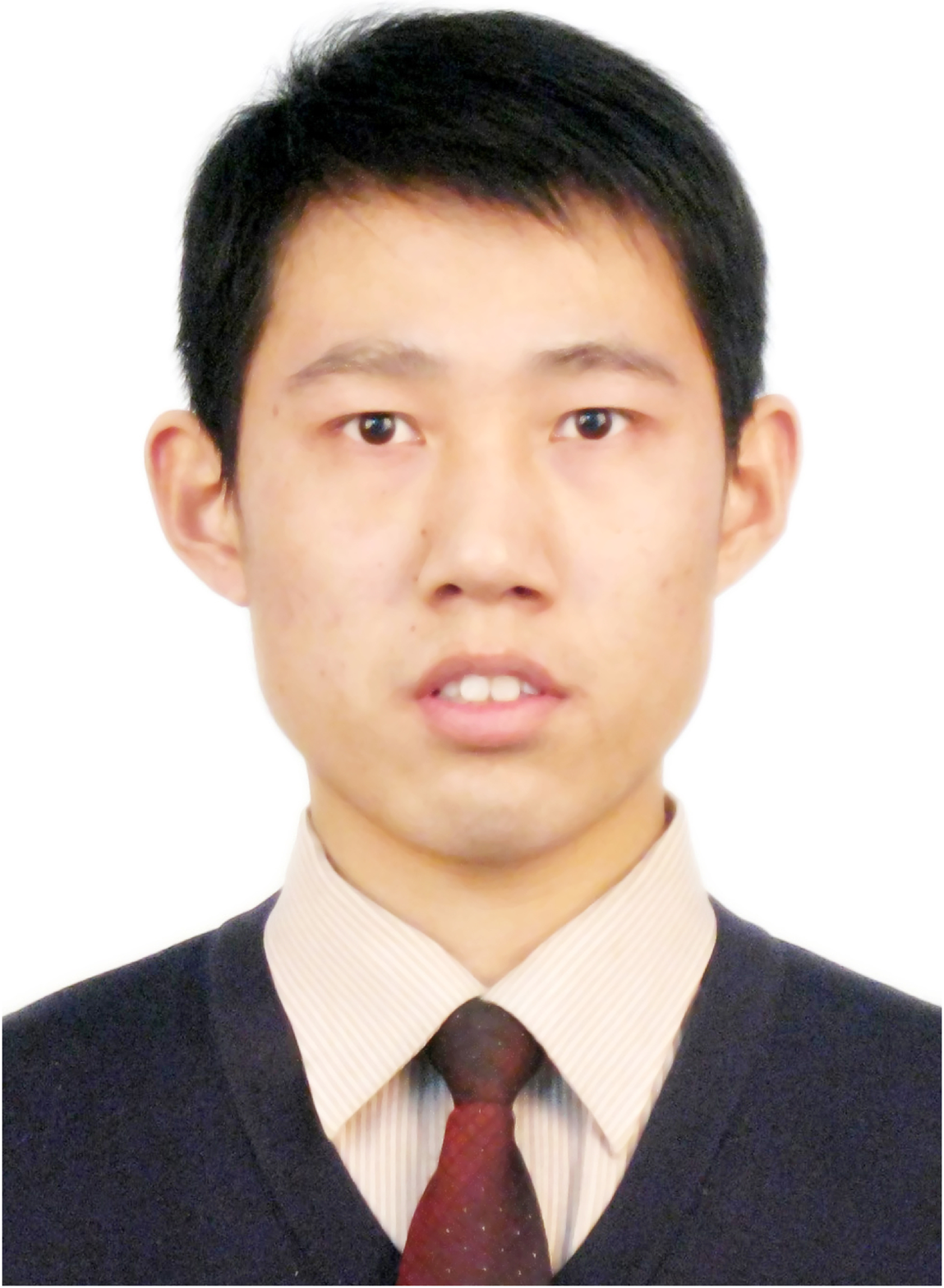}}]{Zhi LIU}
	(S'11-M'14) received the B.E., in computer science and technology from the University of Science and Technology of China, China, in 2009 and Ph.D. degree in informatics in National Institute of Informatics and The Graduate University for Advanced Studies (Sokendai) Tokyo, Japan. He is currently an Junior Researcher (Assistant Professor) at Waseda University, Tokyo, Japan. He was a JSPS research fellow in National Institute of Informatics and The Graduate University for Advanced Studies (Sokendai) from Apr. 2012 to Nov. 2014. From Oct.2009 to Mar. 2014, he was a research assistant in National Institute of Informatics and The Graduate University for Advanced Studies (Sokendai).
	
	His research interest includes wireless networks, video/image processing and transmission. He was the recipient of the IEEE StreamComm2011 best student paper award, VTC2014-Spring Young Researchers Encouragement Award and 2015 IEICE Young Researchers Award. He is and has been a Guest Editor of Sensors and IEICE Transactions on Information and Systems. He has been serving as the TPC-chair of 2017 IEEE Workshop on Game Theory in Computer Communications (in conjunction with IEEE Consumer Communications and Networking Conference (CCNC17)). He is a member of IEEE and IEICE.
\end{IEEEbiography}

\begin{IEEEbiography}[{\includegraphics[width=1in,height=1.25in,clip,keepaspectratio]{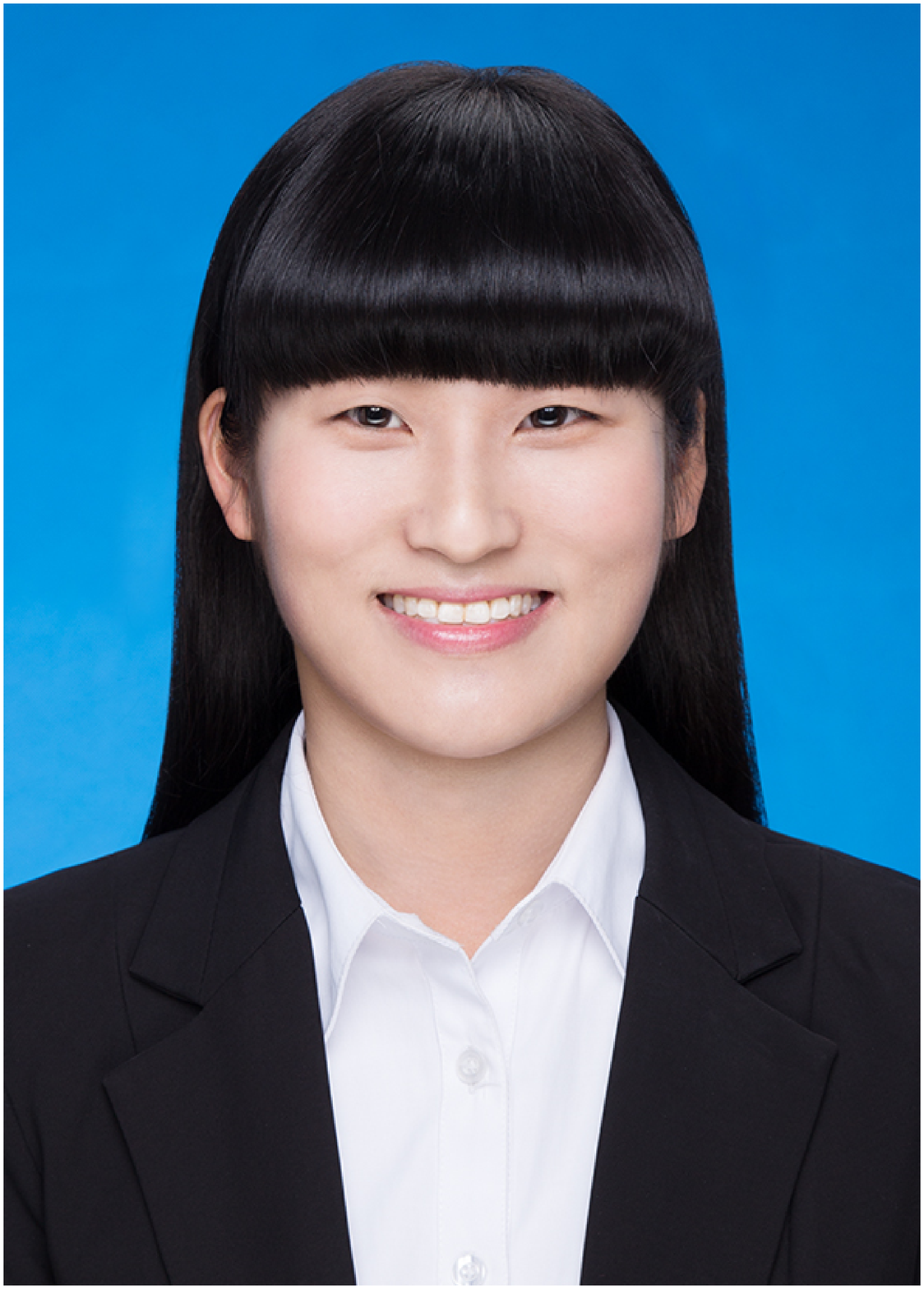}}]{Nana ZHOU}
	received the B.E., in automation from the Hefei University of Technology in 2016. She is currently a postgraduate at Hefei University of Technology, Anhui, China since July 2016. Her research interest includes deep learning and indoor localization.
\end{IEEEbiography}

\begin{IEEEbiography}[{\includegraphics[width=1in,height=1.25in,clip,keepaspectratio]{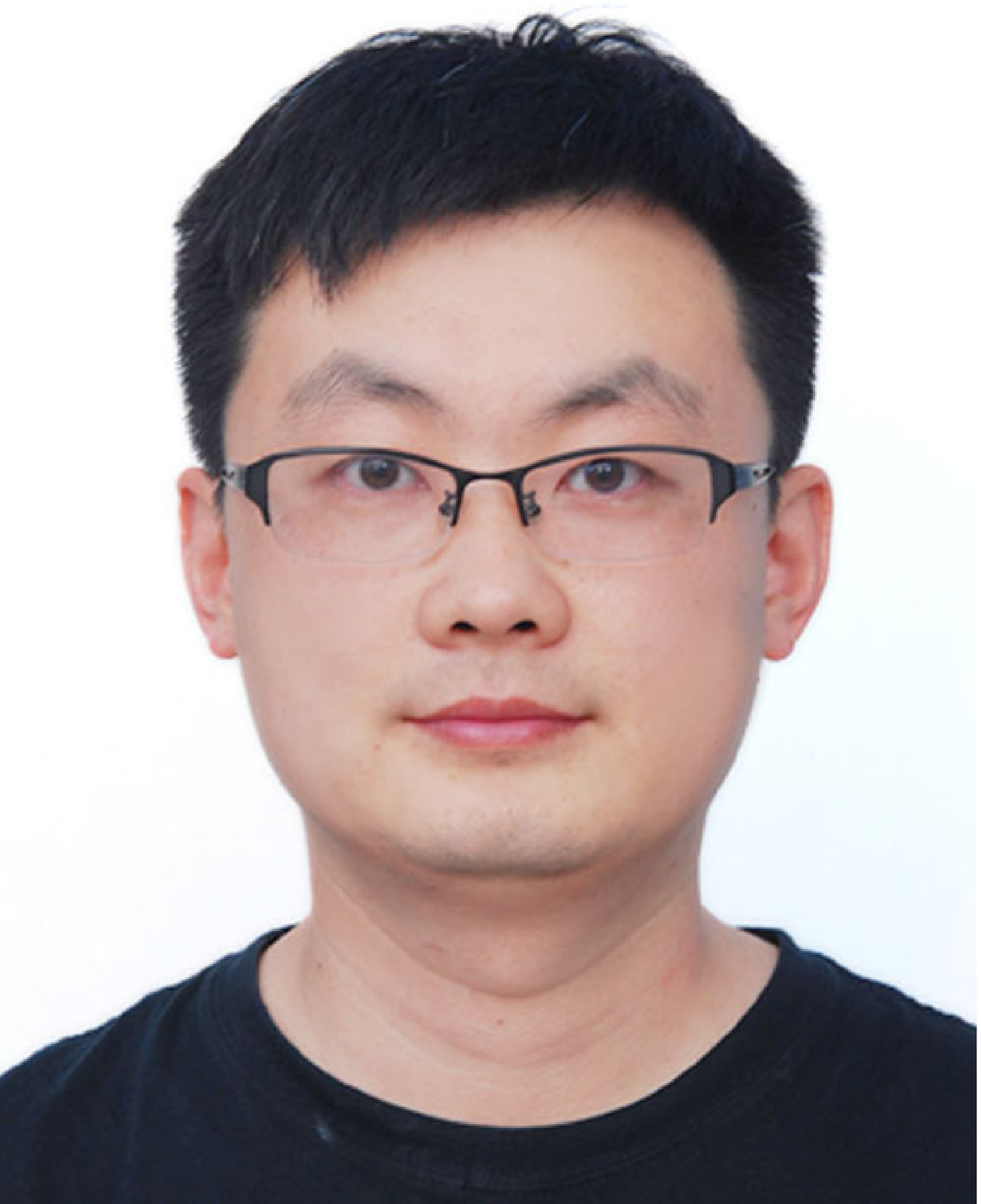}}]{Wei SUN}
	received the B.E. in Automation, the M.E. in Detection Technology and Automatic Equipment and Ph.D. degree in Electrical Engineering from the Hefei University of Technology, in 2004, 2007 and 2012 respectively. He is currently Associate Professor at Hefei University of Technology, Anhui, China. His research interest includes wireless sensor networks and Smart Grid.
\end{IEEEbiography}

\begin{IEEEbiography}[{\includegraphics[width=1in,height=1.25in,clip,keepaspectratio]{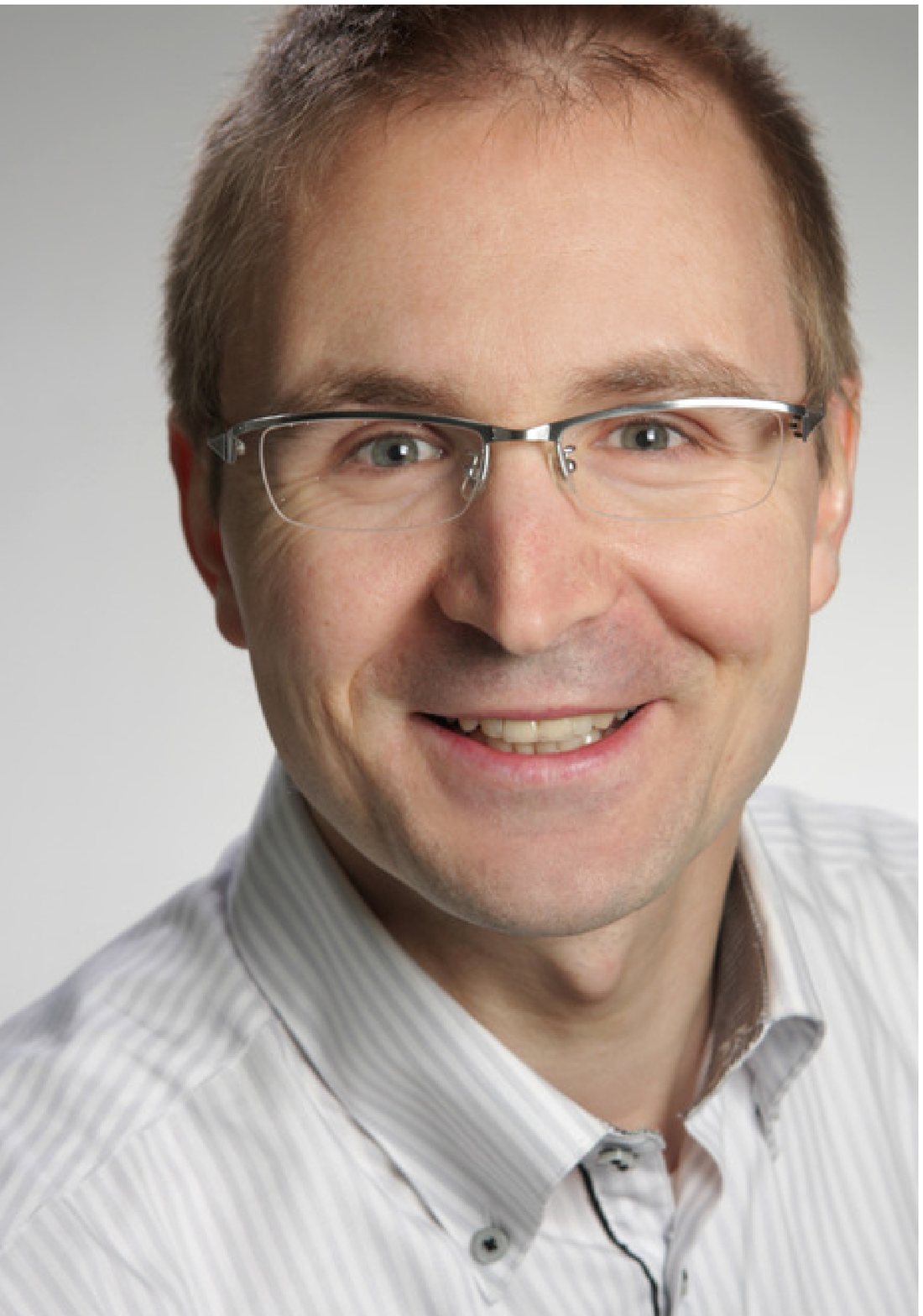}}]{Stephan Sigg}
received his M.Sc. degree in computer science from TU Dortmund, in 2004 and his Ph.D degree from Kassel University, in 2008.
Since 2015 he is an Assistant Professor at Aalto University, Finland.
He has served as a TPC member of numerous conferences including IEEE PerCom, Ubicomp, etc.
His research interests include Usable Security, Pervasive Computing, Activity Recognition, and optimization of randomized algorithms in mobile distributed systems.
\end{IEEEbiography}

\begin{IEEEbiography}[{\includegraphics[width=1in,height=1.25in,clip,keepaspectratio]{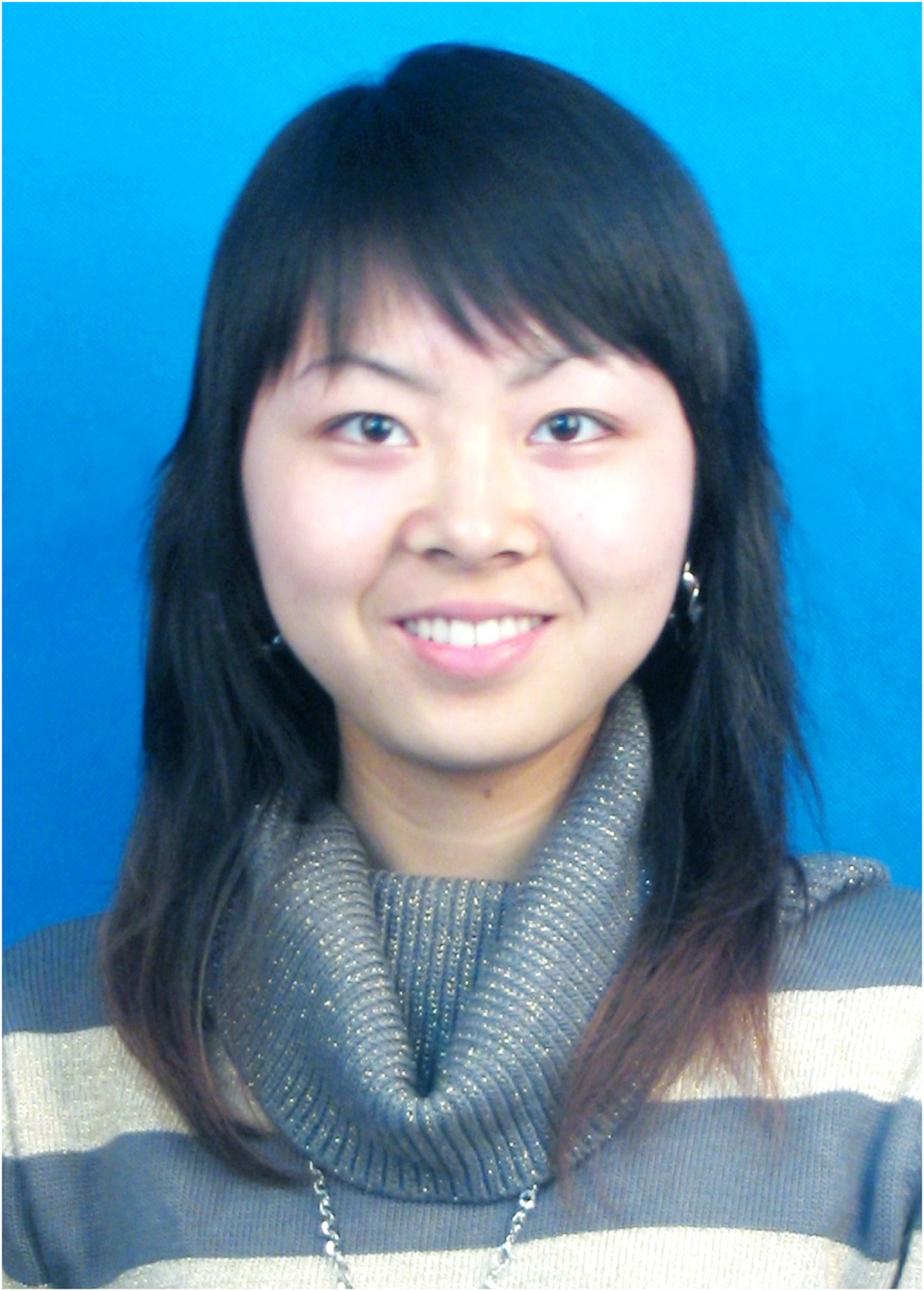}}]{Jie LI}
	received the B.E., in electronic engineering in 2006 and Ph.D. degree in 2011 from University of Science and Technology, China. She is currently an Associate Professor at Hefei University of Technology, Anhui, China. From March 2012 to Sep. 2012, she was a research assistant in National Institute of Informatics, Japan. Her research interest includes cognitive radio network, scalable video multicast and indoor localization using wireless networks.
\end{IEEEbiography}

\end{document}